\documentclass[11pt]{article}
\usepackage[margin=2.5cm]{geometry}
\usepackage[utf8]{inputenc}
\usepackage[mathlines]{lineno}
 
\usepackage{graphicx} 
\usepackage{amsmath}
\usepackage{amsthm}
\usepackage{amssymb}
\usepackage{amstext}
\usepackage{mathtools}
\usepackage{array}   
\usepackage{booktabs} 
\usepackage{multicol}
\usepackage{multirow}
\usepackage{authblk}
\usepackage{nameref,hyperref}
\usepackage{float}
\usepackage{siunitx}
\usepackage[square, numbers]{natbib}

\usepackage{algorithm}
\usepackage{algpseudocode}

\usepackage{dirtytalk}
\MakeRobust{\say}

\usepackage{outlines}

\usepackage[inkscapelatex=false]{svg} 
\usepackage{caption}
\usepackage{subcaption}
\usepackage{orcidlink}
\usepackage{changepage}

\newcolumntype{L}{>{$}l<{$}} %
\newcolumntype{C}{>{\centering$}p{4cm}<{$}}

\newcommand{\comp}[2]{\mathrm{#1}_{\text{#2}}}

\theoremstyle{remark}

\usepackage[normalem]{ulem}
\usepackage{ifthen}

\title{Assessment of Simulation-based Inference Methods for Stochastic Compartmental Models in Epidemiological Research}
\author[1,2,$\dagger$,\orcidlink{0009-0003-9592-8871}]{Vincent Wieland}
\author[1,2,3,$\dagger$,\orcidlink{0009-0000-5023-5404}]{Nils Wa\ss{}muth}
\author[1,\orcidlink{0000-0002-7901-2172}]{Lorenzo Contento}
\author[1,2,3,\orcidlink{0000-0002-0906-6984}]{Martin Kühn}
\author[1,2,$\ast$,\orcidlink{0000-0002-4935-3312}]{Jan Hasenauer}

\affil[1]{Bonn Center for Mathematical Life Sciences, University of Bonn, Bonn, Germany}

\affil[2]{Life and Medical Science Institute, University of Bonn, Bonn, Germany}

\affil[3]{Institute of Software Technology, Department for High-Performance Computing, German Aerospace Center (DLR), Cologne, Germany}

\date{\today}

\affil[$\dagger$]{These authors contributed equally to the work.}
\affil[$\ast$]{To whom correspondence should be adressed; jan.hasenauer@uni-bonn.de.}

\begin{document}

\maketitle

\abstract{
\noindent Global pandemics, such as the recent COVID-19 crisis, highlight the need for stochastic epidemic models that can capture the randomness inherent in the spread of disease. Such models must be accompanied by methods for estimating parameters in order to generate fast nowcasts and short-term forecasts that can inform public health decisions.
This paper presents a comparison of two advanced Bayesian inference methods: 1) pseudo-marginal particle Markov chain Monte Carlo, using an unbiased likelihood estimate obtained by Particle Filter (PF), and 2) Conditional Normalizing Flows (CNF). We investigate their performance on three commonly used compartmental models: A classical Susceptible-Infected-Susceptible (SIS), a Susceptible-Infected-Recovered (SIR) model and a two-variant Susceptible-Exposed-Infected-Recovered (SEIR) model, complemented by an observation model that maps latent trajectories to empirical data. Addressing the challenges of intractable likelihoods for parameter inference in stochastic settings, our analysis highlights how particle-filter-based likelihood estimation and flow-based posterior approximation can provide accurate and robust inference capabilities. The results of our simulation study further underscore the effectiveness of these approaches in capturing the stochastic dynamics of epidemics, providing prediction capabilities for the control of epidemic outbreaks. Results on an Ethiopian cohort study demonstrate operational robustness under real-world noise and irregular data sampling. To facilitate reuse and to enable building pipelines that ultimately contribute to better informed decision making in public health, we make code and synthetic datasets publicly available.}

\clearpage

\section*{Introduction}
The COVID-19 pandemic underscored the critical importance of epidemic modeling research. Rapid, data-driven approaches have demonstrated their capacity to provide public health decision-makers with essential insights for mitigating emerging infectious disease outbreaks and controlling epidemic dynamics~\cite{ethiopian_data, doms2018assessing}. Beyond COVID-19, other outbreaks such as Ebola and MPOX have likewise been designated as \say{public health emergencies of international concern} by the World Health Organization (WHO)~\cite{gpmb2024}. These crises and consequent public health challenges highlight the need for advanced modeling techniques capable of capturing the intricate dynamics of disease spread.

Epidemiological and medical research revealed that the spread of infectious diseases is shaped by a variety of processes, including transmission dynamics, patient responses to treatment, and pathogen mutation. A wide spectrum of mathematical modeling approaches is used to describe the underlying dynamics, spanning from standard ordinary differential equation (ODE) models~\cite{McKendrick_Kermack, hethcote2000} to agent-based models~\cite{bruch2015agent}. On the one hand, ODE-based population models provide a fast to simulate framework, but fall short to describe the variability of events and the resulting complex, stochastic interactions, due to their deterministic nature. Therefore, stochastic models are essential for realistically representing infectious disease dynamics observed in real-world settings, especially in small populations or during the early stages of an outbreak~\cite{britton2010stochastic}. On the other hand, purely stochastic models, such as agent-based and continuous-time Markov chain (CTMC) models~\cite{dragatz_book, infect_dis_modelling} provide highly detailed stochastic representations of individual-level processes, but come at the cost of prohibitive computational burden \cite{petzold_2013}. Mathematical models based on stochastic differential equations -- so called stochastic (meta-)population models -- can solve the trade-off between fast model simulations and keeping the stochasticity in the model~\cite{dragatz_paper}. 
In this study, we consider compartmental models described by stochastic differential equations (SDE). In compartmental models individuals are arranged in a finite number of mutually disjoint compartments and interaction happens through transfer from one compartment to another which is described by transition equations~\cite{Brauer2008}. Their description in terms of SDEs can be seen as a mesoscopic view on the model dynamics. The SDEs are derived as a diffusion approximation to the microscopic, individual-based, discrete-state CTMC description~\cite{dragatz_paper, expansion_vanKampen, fearnhead2014inference} and in the large population limit resemble the macroscopic, population-based, (deterministic) ODE models~\cite{kurtz1970solutions}. For a more detailed discussion on stochastic epidemic modeling see~\cite{infect_dis_modelling, dragatz_paper, AnderssonAndBritton} and references contained therein.

A key challenge in mathematical-epidemiological modeling is determining the unknown model parameters that best explain the observed data. This process, often referred to as parameter inference, is commonly approached by maximizing the likelihood function, which measures the probability of the observed data given specific parameter values. For stochastic models, however, likelihood-based inference is often infeasible when fitting to discrete-time data, as the underlying probability distributions can be highly complex or the parameter space very high-dimensional leading to an intractable likelihood function. In such cases, naive approximations quickly become computationally prohibitive.

Many methods have been developed to address the issue of intractable likelihoods and posterior distributions. Early attempts used data augmentation or employed Markov chain Monte Carlo (MCMC) methods~\cite{dragatz_book, gibson1998estimating, roberts1999bayesian}. Later approaches sought to reduce computational cost through surrogate models such as Gaussian processes~\cite{swallow2022challenges} or linear noise approximations of the underlying Markov jump process~\cite{fintzi2022linear}. Hybrid strategies combining data augmentation with MCMC have also been proposed~\cite{mckinley2014simulation, el2016bayesian}.
Sequential Monte Carlo methods were introduced as a flexible approach for analyzing stochastic state space models and performing Bayesian filtering tasks~\cite{doucet_sequential, chopin2013}. These methods, also known as particle filters (PF), have been used to produce an unbiased likelihood estimate for an MCMC algorithm, leading to pseudo-marginal methods for Bayesian inference on model parameters~\cite{andrieu_roberts, andrieu_2010, chopin_2020}, a popular method for exact Bayesian inference in stochastic models~\cite{chopin_2020, wiqvist_2021}.

Instead of determining model parameters via sampling based on evaluations of the likelihood or an approximation to it, it is possible to conduct the inference procedure solely based on simulations. In the widely used technique of Approximate Bayesian Computing (ABC), simulated and observed data is compared based on summary statistics~\cite{rubin1984, minter_2019}. However, manually selected summary statistics can be uninformative and their selection poses a challenge on its own. Additionally, as these methods require massive numbers of simulations, more recent machine learning approaches have been adapted for simulation-based inference with neural networks~\cite{cranmer2020frontier}. These promising techniques often circumvent hand-crafted summary statistics and directly approximate the distribution of the parameters from samples generated by the simulator. Therein, conditional normalizing flows (CNF) are an advanced method that uses invertible neural networks to generate samples from the posterior distribution of the parameters~\cite{papamakarios2016fast,papamakarios2021normalizing}.

PF and CNFs show great potential for Bayesian inference on stochastic models and are increasingly recognized within the epidemiological modeling community for providing probabilistic estimates of disease characteristics to effectively monitor pandemic outbreaks~\cite{radev2021outbreakflow, storvik2023}.
However, their adoption has been limited by the lack of a comprehensive comparison assessing their relative performance. This work offers a praxis-driven comparison between both inference methods, assessing strengths and weaknesses of the particular approaches, as well as their alignment. 
In the first part, we described the three stochastic compartmental models, the well-known SIS and SIR models~\cite{McKendrick_Kermack} and a two-variant SEIR model, which was used to analyse the spread of COVID-19 in Ethiopia~\cite{ethiopian_data}. Together with the observational model bridging compartment trajectories with empirical data they serve as the foundation for our study. Moreover, we outline Bayesian inference for stochastic differential equation-based models, and explain the chosen techniques of CNF and PF together with the implementation of our comparison workflow. 
In the second part of the work, we outline the results from comparing the two inference methods on synthetically generated data for the chosen compartmental models and additionally using real-data for the two-variant SEIR model. For the two smaller compartmental models, SIS and SIR, we compared the estimated posterior distributions with a reference posterior obtained by Hamiltonian Monte Carlo (HMC) sampling on a discretized version of the stochastic model. Both methods provide robust and reliable inference results on the stochastic versions of the SIS and SIR model with synthetically generated data, validating their implementation at hand.
For the more complex two-variant SEIR model the approximate computation of a reference posterior using HMC methods is not feasible, showcasing the need of the assessed Bayesian inference methods. Analyzing the agreement of the estimated marginal posteriors and model fits given by PF and CNF based on different synthetic datasets, including (partially) missing data, shows that both methods yield good fits to the data at hand. However, ill-conditioning of the model leads to differences in the shape of the marginal posteriors due to difference in parameter space exploration by the inference methods. Applying both methods to an reparametrized version with parameter dimension reduction can overcome these issues and improves posterior alignment. Inference based on the data from~\cite{ethiopian_data} shows the applicability of both methods to real-world data. A discussion of the results, their limitations and further research directions follows in \autoref{section: discussion}.

\section*{Materials and methods}

\subsection*{Stochastic compartmental models}\label{section: models}
Compartmental models are a mathematical framework to describe how individuals in a population move between different states and are fundamental for modeling infectious diseases~\cite{Brauer2008}. Such a model partitions a population into $d_X$ mutually exclusive compartments and is characterized by the vector of compartment sizes $X\in\mathbb{R}^{d_X}$ together with the transitions of individuals between these compartments. 

Let $\mathcal T$ denote the set of admissible transitions. Each transition $(i,j)\in\mathcal T$ is characterized by a stoichiometric vector $s_{i,j}\in\mathbb Z^{d_X}$, which specifies the change in the compartment counts when the transition occurs. For the movement of one individual from compartment $i$ to compartment $j$, $s_{i,j}=e_j-e_i$, where $e_i$ denotes the $i$-th unit vector. The transition intensity is described by a propensity function $a_{i,j}^N:\mathcal S_N\to\mathbb R_+$, defined through
\begin{equation*}
\Pr\bigl(X^N(t+h)=x+s_{i,j}\mid X^N(t)=x\bigr)
= a_{i,j}^N(x)\,h + o(h), \qquad h\downarrow 0.
\end{equation*}
For infection-driven transitions, a common choice is a mass-action propensity of the form
\begin{equation*}
    a_{i,j}^N(X)=c_{i,j}X_i\sum_{k\in I_{i,j}}\frac{X_k}{N},
\end{equation*}
where \(c_{i,j}>0\) is a rate parameter and \(I_{i,j}\) contains the compartments contributing to the force of infection. Other transitions, such as recovery, are typically modeled by linear propensities, for example \(a_{i,j}^N(x)=c_{i,j}x_i\).

The resulting stochastic compartmental model is a continuous-time Markov jump process. In the large-population regime, it can be approximated by the multivariate stochastic differential equation (SDE)~\cite{dragatz_paper}:
\begin{align*}
    dX(t) &= \mu(X(t)) dt +\sigma(X(t))d\mathbf{B}(t)\\
    &= \sum_{(i,j)\in\mathcal{T}}a_{i,j}(X(t))s_{i,j} dt + \sum_{(i,j)\in\mathcal{T}}\sqrt{a_{i,j}(X(t))}s_{i,j}s_{i,j}^T d\mathbf{B}(t) \\
\end{align*}
where $\mathbf{B}$ denotes a vector of independent Brownian motions.
The representation of compartmental models in terms of multivariate SDEs can capture the randomness of disease spread in their diffusion part $\sigma(X(t))$ and retain a drift component $\mu(X(t))$ for representing the deterministic trend of the population~\cite{AnderssonAndBritton, dragatz_book}. This makes them a favorable choice for modeling disease outbreaks and dynamics.

In this work, we consider two examples of SDE-based compartmental models, a simple standard Susceptible-Infected-Recovered (SIR) model~\cite{McKendrick_Kermack} and an extended two-variant Susceptible-Exposed-Infected-Recovered (SEIR) model~\cite{hethcote2000} considering two variants of a single pathogen. 

\begin{figure}[tbp!]
\vspace*{-1cm}
\centering
    \vspace*{-5mm}
    \centering
    \includegraphics[width=0.95\linewidth]{figure_workflow_new.pdf}
    \label{subfig:workflow}
\caption{\textbf{Schematic representations of the models and methods.}\\
(a) Graph of compartments and possible transitions with corresponding rate parameters for the Susceptible - Infected - Susceptible (SIS) model.
(b) Graph of compartments and possible transitions with corresponding rate parameters for the Susceptible - Infected - Recovered (SIR) model. 
(c) Graph of compartments and possible transitions with corresponding rate parameters for the Susceptible-Exposed-Infected-Recovered (SEIR) model with virus variants wild-type (\textbf{wt}) and variant (\textbf{var}).
(d) Workflow for the assessment of the Bayesian inference.}
\label{Fig1}
\end{figure}

\subsubsection*{SIS model}\label{subsec: sis_model}
The simplest compartmental model is the SIS-model~\cite{McKendrick_Kermack} consisting of one \emph{Susceptible} compartment $S$ and one \emph{Infectious} compartment $I$, where people can transition between them (\autoref{Fig1}(a)). This model can be used for infectious diseases that do not confer any long-lasting immunity, such as the common cold or influenza. Following simple mass-action incidence individuals get infected with a rate $\beta \frac{SI}{N}$ and leave the infective class at rate $\gamma I$ (\autoref{tab: description_sis}).

\begin{table}[ht]
    \centering
    \caption{\textbf{Description of the SIS model.} Table including all events, such as initialization and transition of individuals, with their respective rate and model parameters.}
    \label{tab: description_sis}
    \begin{tabular}{||>{\centering\arraybackslash}p{2.8cm}|>{\centering\arraybackslash}p{7cm}|>{\centering\arraybackslash}p{1.8cm}|>{\centering\arraybackslash}p{2cm}||}
        \toprule
        \textbf{Event} & \textbf{Description} & \textbf{Rate} & \textbf{Parameter} \\
        \midrule
            Initialization & Set initial number of infected individuals & - & $\mathrm{I}_0$\\
            Infection & $\comp{S}{} \rightarrow\comp{I}{}$ & $\beta\frac{\mathrm{S}(t)\mathrm{I}(t)}{\mathrm{N}(t)}$ & $\beta$ \\
            Recovery & $\comp{I}{}\rightarrow\comp{S}{}$ & $\gamma \mathrm{I}(t)$ & $\gamma^{-1}$\\
            
        \bottomrule
    \end{tabular}
\end{table}

\subsubsection*{SIR model}\label{subsec: sir_model}
The standard SIR model (\autoref{Fig1}(b)) considers three compartments: $S$ (\emph{Susceptible}), $I$ (\emph{Infectious}) and $R$ (\emph{Recovered}). The model describes the number of individuals in these compartments given the transition parameters $\beta$ (the \emph{transmission rate}) and $\gamma$ (the \emph{recovery rate}).
Due to its simplicity, the solution of the deterministic SIR model and its properties can be analysed analytically~\cite{harko2014exact} and it often serves as an initial model for the modeling of unknown diseases~\cite{daley1999epidemic}. 
For consistency across models, in the following we compare recovery in terms of the mean infectious period, $\gamma^{-1}$, rather than $\gamma$. The events dynamics are described by three compartments and the transitions between them (\autoref{tab: description_sir}).

\begin{table}[ht]
    \centering
    \caption{\textbf{Description of the SIR model.} Table including all events, such as initialization and transition of individuals, with their respective rate and model parameters.}
    \label{tab: description_sir}
    \begin{tabular}{||>{\centering\arraybackslash}p{2.8cm}|>{\centering\arraybackslash}p{7cm}|>{\centering\arraybackslash}p{1.8cm}|>{\centering\arraybackslash}p{2cm}||}
        \toprule
        \textbf{Event} & \textbf{Description} & \textbf{Rate} & \textbf{Parameter} \\
        \midrule
            Initialization & Set initial number of infected individuals & - & $\mathrm{I}_0$\\
            Infection & $\comp{S}{} \rightarrow\comp{I}{}$ & $\beta\frac{\mathrm{S}(t)\mathrm{I}(t)}{\mathrm{N}(t)}$ & $\beta$ \\
            Recovery & $\comp{I}{}\rightarrow\comp{R}{}$ & $\gamma \mathrm{I}(t)$ & $\gamma^{-1}$\\
        \bottomrule
    \end{tabular}
\end{table}

\subsubsection*{Two-variant SEIR model}\label{subsec: seir_model}
Many extensions of the SIR model have been developed to account for the heterogeneity of various infectious diseases~\cite{keeling2008}, for example the Susceptible-Exposed-Infectious-Recovered model, in which the Exposed compartment covers the latent period between being infected and becoming infectious.
Recently, an extension to the ODE-based SEIR model was introduced to capture the early COVID-19 dynamics in Ethiopia~\cite{ethiopian_data}. In addition to a classical SEIR model, this model distinguishes between two virus variants, namely the wild-type and a novel variant, leading to a total of 10 compartments (\autoref{Fig1}(c)). 

The novel variant is assumed to have a 35\% longer infectious period, following the original formulation of the two‑variant SEIR model in \cite{ethiopian_data}. We adopt this fixed factor to remain consistent with the published model on which our comparison is based. Additionally, immunity from the wild-type variant does not protect against reinfection with the novel variant. However, an infection with the novel variant type confers immunity against subsequent infections with both the wild-type and the novel variant. The dynamics of the compartment sizes are described by initialization, variant entering and transitions between the compartments (\autoref{tab: description_seir2v}).

The model exhibits non-identifiability issues in the sense that different parameter combinations, lying along certain curves, can generate nearly indistinguishable epidemic trajectories (see Supplementary Information~\nameref{supp:file1}). Therefore, in addition to this model, we also consider a reparametrized version of the model with a reduced number of model parameters to be estimated. For this, we fix $\kappa^{-1}=5$, since $\kappa$ seems to participate in the non-identifiable parameter combinations and use the following redefined parameters
\begin{equation}\label{eq: reparam}
    r_0 = \beta\cdot\gamma^{-1}, \quad
    e_0 = \kappa^{-1} + \gamma^{-1}, \quad
    s_0 = s\cdot\gamma^{-1},
\end{equation}
while keeping $t_{\mathrm{var}}$ and $I_{0}$ as well as the observational and noise models unchanged. This transformation allows a comparisons of the inference methods CNF and PF in either parameter space without altering the underlying system.

\begin{table}[ht]
    \centering
    \small
    \caption{\textbf{Description of the two-variant SEIR model.} Table including all events, such as initialization and transition of individuals, with their respective rate and model parameters.}
    \label{tab: description_seir2v}
    \begin{tabular}{||>{\centering\arraybackslash}p{3cm}|>{\centering\arraybackslash}p{6.5cm}|>{\centering\arraybackslash}p{3.2cm}|>{\centering\arraybackslash}p{1.8cm}||}
        \toprule
        \textbf{Event} & \textbf{Description} & \textbf{Rate} & \textbf{Parameter} \\
        \midrule
            Initialization & Set initial number of infected individuals & - & $\mathrm{I}_0$\\
            Variant entry & 500 individuals infected with variant enter the population & - & $t_\mathrm{var}$\\
            Wild-type infection & $\comp{S}{} \rightarrow\comp{E}{wt}$ & $\beta\frac{\mathrm{S}(t)\mathrm{I_{wt}}(t)}{\mathrm{N}(t)}$ & $\beta$ \\
            Wild-type incubation & $\comp{E}{wt}\rightarrow\comp{I}{wt}$ & $\kappa \mathrm{E_{wt}}(t)$ & $\kappa^{-1}$\\
            Wild-type recovery & $\comp{I}{wt}\rightarrow\comp{R}{wt}$ & $\gamma \mathrm{R_{wt}}(t)$ & $\gamma^{-1}$ \\
            Variant infection & $\comp{S}{}\rightarrow\comp{E}{var}$ & $\beta\frac{\mathrm{S}(t)(\mathrm{I_{var}}(t)+\comp{I}{wt·var}(t))}{\mathrm{N}(t)}$ & $\beta$ \\
            Variant incubation & $\comp{E}{var}\rightarrow\comp{I}{var}$ & $\kappa \mathrm{E_{var}}(t)$ & $\kappa^{-1}$ \\
            Variant recovery & $\comp{I}{var}\rightarrow\comp{R}{var}$ & $\frac{\gamma}{1.35}\comp{I}{var}(t)$ & $\gamma^{-1}$ \\
            Second infection & $\comp{R}{wt}\rightarrow\comp{E}{wt·var}$ & $\beta\frac{\mathrm{R_{wt}}(t)(\mathrm{I_{var}}(t)+\comp{I}{wt·var}(t))}{\mathrm{N}(t)}$ & $\beta$ \\
            Second incubation & $\comp{E}{wt·var}\rightarrow\comp{I}{wt·var}$ & $\kappa \comp{E}{wt·var}(t)$ & $\kappa^{-1}$ \\
            Second recovery & $\comp{I}{wt·var}\rightarrow\comp{R}{wt·var}$ & $\frac{\gamma}{1.35}\comp{I}{wt·var}(t)$ & $\gamma^{-1}$ \\
        \bottomrule
    \end{tabular}
\end{table}

\subsubsection*{Observation model}\label{subsec: observation model}
The course of an epidemic is usually only partially observed and data is often only available in aggregated form on population level. For all models introduced above, we consider an observation model reflecting one to two often measured quantities to monitor infectious diseases. For the SIS model, the observable is
\begin{align}\label{eq: sis_observation_model}
        y(t)=\left(\begin{array}{c}
        \frac{\comp{I}{}(t)}{\comp{N}{}(t)}
    \end{array}\right).
\end{align}
For the SIR model, the observables are
\begin{align}\label{eq: sir_observation_model}
        y(t)=\left(\begin{array}{c}
        \frac{\comp{I}{}(t)}{\comp{N}{}(t)}\\[0.7em]
        \frac{\comp{N}{}(t)-\comp{S}{}(t)}{\comp{N}{}(t)}
    \end{array}\right).
\end{align}
For the two-variant SEIR model, they are given by
\begin{align}\label{eq: seir_observation_model}
    y(t)=\left(\begin{array}{c}
        \frac{s(\comp{I}{wt}(t)+\comp{I}{wt·var}(t)+\comp{I}{var}(t))}{\comp{N}{}(t)}\\[0.7em]\frac{\comp{R}{wt}(t)+\comp{E}{wt·var}(t)+\comp{I}{wt·var}(t)+\comp{R}{wt·var}(t)+\comp{R}{var}(t)}{\comp{N}{}(t)}
        \\
    \end{array}\right),
\end{align}
where the first entry represents the latent prevalence of currently infected individuals. In practice, this quantity is not observed directly but is inferred from testing data. One method of inference is to consider the proportion of positive tests among all tests performed, under the assumption that the cohort tested is representative of the population as a whole. The observation model therefore treats testing-derived prevalence as noisy observations of the underlying population prevalence given by the model. The second entry represents seroprevalence, i.e., the proportion of individuals with antibodies due to a prior infection. Because the observation process may have several limitations, such as incomplete reporting or under-ascertainment, an additional scaling parameter $s$ is used to account for the fact that only a fraction of infections may be captured in the reported testing data~\cite{ethiopian_data}. 

To reflect measurement uncertainty, synthetic realizations of the observables are perturbed with noise. In this study, we leverage additive Gaussian noise. That means we have
\begin{align*}
    \overline{y}_1(t_i) &= y_1(t_i)+\varepsilon_{i,1}, \qquad \varepsilon_{i,1}\sim\mathcal{N}(0,\sigma_{i,1}^2),\\
    \overline{y}_2(t_i) &= y_2(t_i)+\varepsilon_{i,2}, \qquad \varepsilon_{i,2}\sim\mathcal{N}(0,\sigma_{i,2}^2),
\end{align*}
where we truncate $\overline{y}_{1}(t_i)$ and $\overline{y}_{2}(t_i)$ at a lower bound of 0 and at an upper bound of 1 to avoid unrealistic observations. As in the dataset given by~\cite{ethiopian_data}, the standard deviations $\sigma_{i,1}, \sigma_{i,2}$ were computed directly using a normal approximation to the binomially distributed testing count data. Alternative probability distributions may be more appropriate in other contexts.

From each of the stochastic compartmental models we generated synthetic data emulating the time course of a disease outbreak. We generated dense datasets with observations every seven or ten days. We used two types of parameter regimes: 
\begin{enumerate}
    \item hand‑selected values chosen to represent plausible epidemiological regimes, and 
    \item parameters sampled directly from the corresponding priors as an unbiased baseline
\end{enumerate} Prior draws that generated trajectories with collapsed variance were discarded and resampled to maintain numerical stability of the particle filter. Additionally, to create realistic, heterogeneous observation schedules, we randomly subsampled the dense time series into sparse datasets reducing the number of time points from 57 to fewer than 15 per observable. We also introduced partial missingness in the sparse datasets, where only one of the two observables was available at certain time points. Details on the computation of measurement noise and data generation are provided in the Supplementary Information~\nameref{supp:file1}.

\subsection*{Bayesian inference for stochastic compartmental models}\label{section: inference methods}
To infer the unknown model parameters $\mathbf{\theta}$ given data ${\mathcal{D}^*=\{\overline{y}_{t_i}\mid i\in\{1,...,n\}\}}$, we apply Bayesian inference techniques. For this, we denote the space of unobserved states of the diffusion process by $\mathcal{X}$.

Bayes' rule allows to combine prior information $p(\theta)$ (e.g., from expert judgment, previous studies, biological constraints, or model restrictions) with the likelihood $p(\mathcal{D}^*\mid\theta)$, which measures how probable the observed data are given specific parameter values, yielding the posterior distribution
\begin{align}\label{eq: posterior}
p(\theta \mid\mathcal{D}^*)=\frac{p(\mathcal{D}^*\mid\theta)p(\theta)}{p(\mathcal{D}^*)}\,\propto \,p(\mathcal{D}^*\mid\theta)p(\theta).
\end{align}
The posterior represents the updated belief about the parameters after taking the observed data into account via the likelihood, effectively integrating both previous knowledge and the new information provided by the data.  While the full posterior distribution is typically of primary interest, we additionally consider the mode (called \emph{maximum a-posteriori}) of the posterior distribution as the best point estimate~\cite{bernardo1994bayesian}. 

One challenge in applying Bayesian inference to stochastic compartmental models is dealing with an intractable posterior distribution, because evaluating the likelihood 
\begin{align}\label{eq: likelihood}
    p(\mathcal{D}^*\mid\theta)=\int_{\mathcal{X}}p(X, \mathcal{D}^*\mid\theta)dX
\end{align}
requires marginalizing over stochastic realizations of the unobserved states at measurement time points.

In most cases, evaluating the density of hidden states analytically is impossible~\cite{Doucet_SMC}, which prohibits evaluating the likelihood. Even in the case of a tractable likelihood, normalization of the posterior is done via the marginal likelihood, which can remain intractable. High‑dimensional parameter spaces further make discretization or numerical integration prohibitively expensive~\cite{wilkinson2010, hasenauer2013thesis}.

Approximation techniques that estimate the posterior distribution without directly computing the complicated likelihood function enable robust parameter estimation, even in settings where traditional methods may fail due to mathematical or computational complexity. In this section, we will introduce two approaches (\textit{particle filters} (PF)~\cite{golightly_particle_2011} and one particular class of conditional density estimators, \textit{conditional normalizing flows} (CNF)~\cite{papamakarios2016fast, cranmer2020frontier}) that circumvent the computation of the likelihood function. PF approximate the likelihood with an unbiased estimator obtained by sequential importance sampling, and conditional normalizing flows learn a series of invertible mappings to transform a Gaussian density into the desired density.

\subsubsection*{Particle Markov chain Monte Carlo methods}
Particle Markov chain Monte Carlo (PMCMC) methods are based on the idea of using an unbiased estimator of the likelihood function, obtained by a particle filter, inside a Markov chain Monte Carlo (MCMC) method~\cite{golightly_particle_2011}.

Particle filters are sequential Monte Carlo (SMC) techniques based on importance sampling with resampling used for handling non-linear and non-Gaussian time-series models. Let $X_{t_0:t_M}$ denote a discrete time-series of the unobserved states and $y_{t_0:t_M}$ a time-series of observations at the same time points. The main idea is to iterate over the following three steps. First, $N$ particles, i.e., realizations of the dynamic process, are propagated through the latent space over time. This means that, given the trajectory of the latent process for the $k$-th particle and the observables until time $t_{i-1}$, we sample the state of the latent process at time $t_i$, $X_{t_i}^{(k)}\sim q_{t_i}\left(\cdot\mid X_{t_0:t_{i-1}}^{(k)}, y_{t_0:t_{i-1}},\theta\right)$, based on a chosen proposal distribution $q_{t_j}$. Second, at each time point, particles are weighted, based on their agreement with the observed values for that time point. And third, they are resampled using auxiliary variables $A_{t_j}^{1:N}$ to discard low-weight particles. In practice, the resampling is not done in every iteration, but only if the variability of the weights becomes too large, as measured by the effective sample size (ESS) (for more details see Supplementary Information \nameref{supp:file1}). The resampled particles are then again propagated forward to the next time point. Hence, for each time $t_i$, the particle filter constructs a system of $N$ weighted particles $\{X_{t_i}^{(k)},\, w_{t_i}^{(k)}\}_{k=1}^N$ that approximates the filtering distribution $p(X_{t_i}\,\mid\,y_{t_0:t_i},\theta)$. Mathematically, this is closely linked to the framework of Feynman-Kac models~\cite{moral2004feynman, andrieu_2010, chopin_2020}.

For the task of estimating the likelihood in stochastic compartmental models, we use a \textit{bootstrap filter} Algorithm~\ref{algorithm:bootstrap_filter}. This is a PF algorithm, where particles at the next time point are generated solely based on the transition probabilities $p_\theta(X_{t_j}\mid X_{t_{j-1}})$ of the latent process and then those new particles are filtered based on their agreement with the data at hand $g_\theta(y_{t_j}\mid X_{t_j})$, where the function $g_\theta$ is determined by the observation and noise model.
The bootstrap filter is easy to implement and widely applicable due to its simplicity. It only requires evaluation of the observation model $g_\theta(y_{t_j}\mid X_{t_j})$. However, it is not necessary to compute the transition densities $p_\theta(X_{t_j}\mid X_{t_{j-1}})$; we solely need the ability to simulate the state of the process at the next time point given the current state, making it practical for any complex or stochastic model.

\begin{algorithm}[ht]
\caption{Bootstrap Filter with Adaptive Resampling}
\label{algorithm:bootstrap_filter}
\begin{algorithmic}[1]
\Require parameter $\theta$, number of particles $N$, observation density function $g_{\theta}$, transition density function $p_{\theta}$, initial distribution $\pi$\\
\Comment{Operations involving index $n$ must be performed for $n=1,...N$}
\State Sample $X_{t_0}^{(n)}\sim \pi(x_{t_0})$
\State $w_{t_0}^{(n)} \leftarrow g_\theta(y_{t_0}\mid X_{t_0}^{(n)})$
\State $W_{t_0}^{(n)} \leftarrow \frac{w_{t_0}^{(n)}}{\sum_{k=1}^N w_{t_0}^{(k)}}$
\For{$i=1$ to $M$}
    \If{$\mathrm{ESS(W}_{t_{i-1}}^{1:N})<\mathrm{ESS_{min}}$}
        \State Sample $A^{1:N}_{t_i}\sim \text{resample}(W_{t_{i-1}}^{1:N})$ \hspace{0.5cm}\Comment{ See Supplementary Information~\nameref{supp:file1}}
        \State$\hat{w}_{t_{i-1}}^{(n)} \leftarrow 1$
    \Else
        \State $A_{t_i}^{(n)} \leftarrow n$ \hspace{3.7cm}\Comment{ No resampling}
        \State $\hat{w}_{t_{i-1}}^{(n)} \leftarrow w_{t_{i-1}}^{(n)}$
    \EndIf
    \State Sample $X_{t_i}^{(n)}\sim p_\theta(x_{t_i}\mid X_{t_{i-1}}^{A_{t_j}^{(n)}})$
    \State $w_{t_i}^{(n)} \leftarrow \hat{w}_{t_{i-1}}^{(n)}$ $g_\theta(y_{t_i}\mid X_{t_{i}})$
    \State $W_{t_i}^{(n)} \leftarrow \frac{w_{t_i}^{(n)}}{\sum_{k=1}^N w_{t_i}^{(k)}}$
\EndFor
\end{algorithmic}
\end{algorithm}

Although the name \emph{filtering} refers to the task of estimating the latent state $X_{t_i}$ online given data $y_{t_0:t_i}$, a PF yields, as a by-product, a scheme to compute an unbiased estimator of the likelihood contributions ${\{p(y_{t_0:t_i}\,\mid\,\theta)\}}_{i \geq 0}$, crucial for the task of parameter inference~\cite{chopin_2020}. This estimator is obtained by computing the normalizing constants
\begin{align}
    l_{t_i}^N(\theta)=\frac{\sum_{k=1}^N w_{t_i}^{(k)}}{\sum_{k=1}^Nw_{t_{i-1}}^{(k)}}\quad\text{and}\quad L_{t_i}^N(\theta)=\prod_{j=0}^il_{t_j}^N,
\end{align}
where $L_{t_i}^N(\theta)$ corresponds to the desired estimate of $p(y_{t_0:t_i}\mid\theta)$. The estimate $L_{t_M}^N(\theta)$ can then be used inside a pseudo-marginal MCMC algorithm to sample from the posterior $p(\theta \mid y_{t_0:t_M})$.

In this work, we use the likelihood estimate obtained by the bootstrap filter within the acceptance probability of a Metropolis-Hasting (MH) algorithm as the outer MCMC scheme, making it a member of the class of Pseudo-Marginal Metropolis-Hasting methods (PMMH)~\cite{andrieu_roberts, andrieu_2010, chopin_2020}. The full PMMH-algorithm is provided in the Supplementary Information~\nameref{supp:file1}.

Using an unbiased estimator $\hat{p}(y|z, \theta)$ of the likelihood $p(y|\theta)$ that is obtained by the use of auxiliary random variables $z\sim r(\cdot)$ within the acceptance of a MH algorithm then corresponds to simulating a Markov chain $\{\theta^i, z^i\}_{i\geq 0}$ targeting the joint density
$$
\bar{p}(\theta, z)\propto p(\theta)r(z)\hat{p}(y\mid \theta, z).
$$
For the bootstrap filter, we can, for instance, think of the $A_{t_j}^{1:N}$ used in the resampling as being the auxiliary random variables. The joint density admits the correct posterior density of the model $p(\theta\mid y)$ as a marginal and therefore the algorithm samples the parameter vectors $\theta$ from the true posterior~\cite{andrieu_roberts, andrieu_2010, doucet_2015}. With respect to an extended distribution, PMMH algorithms are standard MCMC samplers.

One can use any unbiased estimator within any MCMC algorithm to produce a PMCMC algorithm. Different filtering and sampling algorithms impose different practical and implementational challenges and we focus here on the most general applicable version~\cite{doucet_2015, chopin_2020}. In the subsequent sections \textit{PF method} always refers to a PMMH algorithm consisting of a bootstrap filter inside an adaptive Metropolis-Hastings algorithm.

\subsubsection*{Neural posterior estimation and conditional normalizing flows}\label{sec:abi_cnf}
Neural posterior estimation (NPE) is a simulation-based inference (SBI) method which frames posterior construction as a (conditional) density-estimation problem  \cite{papamakarios2016fast}. Given a simulator that produces samples $(\mathcal{D},\theta)$ by first drawing $\theta\sim p(\theta)$ and then $\mathcal{D}\sim p(\mathcal{D}\mid \theta)$, NPE fits a conditional neural network $p_{\phi}(\theta\mid\mathcal{D})$ to approximate the true posterior $p(\theta\mid\mathcal{D})$. Here $p(\mathcal{D}\mid \theta)$ denotes the implicit likelihood defined by the stochastic forward model. It is the distribution over datasets induced by running the forward model at a fixed parameter $\theta$. NPE sidesteps exact likelihood evaluation by relying only on these simulated samples.

Typically the data $\mathcal{D}$ are preprocessed by a (learned) summary statistic~\cite{radev2020bayesflow}, which serves as the input to the conditional density estimator. For clarity, this preprocessing step is omitted from the notation below.

In this work, we focus on Conditional Normalizing Flows (CNFs)~\cite{rezende2015variational}. Normalizing flows have since been substantially developed and systematized
\cite{papamakarios2021normalizing, BoeltsDeistler_sbi_2025, radev2023bayesflow}.
In general, normalizing flows transform a simple base distribution (e.g., a standard Gaussian) into a complex target distribution via a sequence of invertible, differentiable mappings. In our setting, this means that a normalizing flow defines a diffeomorphism $f_{\mathcal{D}^*}$ such that (for fixed $\mathcal{D}^*$) the posterior $p(\theta\mid\mathcal{D}^*)$ is the \say{pushforward} of the base distribution $q$ along $f_{\mathcal{D}^*}$, i.e.,
\begin{align*}
    \theta \sim p(\theta\mid\mathcal{D}^*) \Leftrightarrow \theta= f_{\mathcal{D}^*}^{-1}(z), \quad z \sim q(z),
\end{align*}
This construction enables efficient sampling and exact density evaluation via the change‑\allowbreak of‑\allowbreak variables formula, making normalizing flows a powerful tool for modeling complex probability distributions.

CNFs extend this idea by allowing the transformation to additionally depend on datasets $\mathcal{D}$, generated by first drawing $\theta \sim p(\theta)$ and then $\mathcal{D} \sim p(\mathcal{D}\mid \theta)$. Rather than learning a single transformation $f$, CNFs learn a family of conditional flows $\{f_{\mathcal{D}}\}$, with each $f_{\mathcal{D}}$ mapping the base distribution $q$ to the posterior $p(\theta \mid \mathcal{D})$~\cite{winkler2019conditional}. This conditioning enables CNFs to capture how the posterior changes with different observations, making them particularly well-suited for amortized inference tasks, where a model learns to perform inference efficiently across multiple datasets rather than solving each case from scratch~\cite{kobyzev2020normalizing, rezende2015variational}.

Instead of hand-designing each $f_{\mathcal{D}}$, we parameterize the family using a neural network $f_{\phi}(\cdot,\,\mathcal{D})$
whose parameters $\phi$ are shared across all $\mathcal{D}$, but whose behavior is modulated by $\mathcal{D}$ itself~\cite{ardizzone2019guided}.
We train the neural network by minimizing the Kullback-Leibler divergence between the true and learned posteriors for all $\mathcal{D}$ in the training distribution~\cite{weilbach2020structured,siahkoohi2020conditional,radev2023bayesflow}. This means, our optimization objective is finding $\hat{\phi}$
\begin{align*}
    \displaystyle \hat{\phi}=\arg\min_{\phi}\; -\mathbb{E}_{(\theta,\mathcal{D})\sim p(\theta)p(\mathcal{D}\mid\theta)}\big[\log p_{\phi}(\theta\mid\mathcal{D})\big]
\end{align*}
For CNFs this requires computing the change‑of‑variables log‑density
\begin{align*}
    \log p_{\phi}(\theta\mid \mathcal{D})=\log q(f_{\phi}(\theta,\mathcal{D}))+\log\Big|\det\Big(\frac{\partial f_{\phi}(\theta,\mathcal{D})}{\partial \theta}\Big)\Big|
\end{align*}
for simulator samples and optimizing $\phi$ by stochastic gradient descent. This amortizes the cost of inference: Once trained, the CNF can quickly return an approximate posterior for any new $\mathcal{D}$ within the support of the training distribution. Algorithm \ref{alg:amortized-cnf} summarizes the amortized training loop and the inference procedure for drawing approximate posterior samples via inverse flow sampling.

\begin{algorithm}
\caption{Amortized training and inference for conditional normalizing flows}
\label{alg:amortized-cnf}
\begin{algorithmic}[1]
\Require Data source $S$ (pre‑simulated dataset or simulator callable $\sim p(\theta)p(\mathcal{D}\mid\theta)$), base density $q(z)$, conditional flow $f_{\phi}(\cdot;\mathcal{D})$, optimizer Opt, learning rate $\eta$, batch size $N$, inference data $\mathcal{D}^*$, inference sample count $M$
\Ensure Trained parameters $\phi$; sampler for approximate posterior $p_{\phi}(\theta\mid\mathcal{D}_*)$

\algorithmicrepeat
  \State $\{(\theta_i,\mathcal{D}_i)\}_{i=1}^N \leftarrow \text{sample from $S$ (dataset or simulator)}$
  \For{$i=1$ to $N$}
    \State $z_i \leftarrow f_{\phi}(\theta_i;\mathcal{D}_i)$
    \State $\log p_{\phi}(\theta_i\mid\mathcal{D}_i) \leftarrow \log q(z_i) + \log\Big|\det\Big(\frac{\partial f_{\phi}(\theta_i;\mathcal{D}_i)}{\partial \theta}\Big)\Big|$
  \EndFor
  \State $\mathcal{L} \leftarrow -\frac{1}{N}\sum_{i=1}^N \log p_{\phi}(\theta_i\mid\mathcal{D}_i)$
  \State $\phi \leftarrow \text{Opt}(\phi,\nabla_{\phi}\mathcal{L},\eta)$
\algorithmicuntil{convergence of $\phi$}

\Statex
\Procedure{Infer}{$\mathcal{D}_*$, $M$} \Comment{$M$ is number of posterior samples}
  \For{$m=1$ to $M$}
    \State Sample $z_m \sim q(z)$
    \State $\theta_m \leftarrow f_{\phi}^{-1}(z_m;\mathcal{D}_*)$
  \EndFor
  \State \Return $\{\theta_m\}_{m=1}^M$ 
\EndProcedure
\end{algorithmic}
\end{algorithm}

This amortization not only enables fast inference once training is complete, but also facilitates simulation‑based calibration (SBC) diagnostics by allowing posterior approximations to be generated efficiently across many simulated datasets. At the same time, performance of CNFs is tied to the support of the training distribution: the model can interpolate to new values if they are close to the training support, but performance degrades when extrapolating far beyond it~\cite{winkler2019conditional,talts2018validating}. In practice, this amortized structure is precisely what makes SBC feasible at scale, since once trained, the CNF can be evaluated repeatedly on equally structured datasets without retraining.

\subsection*{Implementation of the methods and their comparison}\label{section: Implementation}
In order to render the simulation based approaches computationally feasible, model simulations were conducted using the efficiency of the Julia programming language and the flexible SDE modeling capabilities of \texttt{SciML} and \texttt{ModelingToolkit.jl}~\cite{julia_stochdiffeq, ma2021modelingtoolkit}. To approximate the solution of the SDE, we implemented an Euler-Maruyama scheme that constrains trajectories to stay non-negative, effectively clipping potential negative parts and setting them to zero.

For PF, existing software packages were unsuitable as they either lacked Julia support or restricted model specification to simulator-only interfaces. We therefore implemented a custom particle filter in Julia, drawing inspiration from the Python package \texttt{Particles}~\cite{chopin_2020}. As an outer sampling scheme, we employed an adaptive Metropolis–Hastings algorithm~\cite{adaptive_mh}, using the implementation in \texttt{pyPESTO}~\cite{pypesto}, to account for potential multimodality in the posterior distribution. Missing data were handled by setting the likelihood contribution of the corresponding observation to zero. 
To mitigate particle degeneracy, we set the number of particles to 200, ensuring that the variance of the likelihood estimate remained below one for parameters around the posterior mode (Supplementary Information~\nameref{supp:file1}, Fig. S1.1). For the outer MH algorithm, we ran 4 Markov chains with a length of 50,000 samples each. We then discarded a burn-in period of the first 25,000 samples per chain. After that, convergence was evaluated using rank‑normalized Gelman–Rubin statistics ($\hat{R}$) and the ESS with autocorrelations up to a maximal lag of 250, both from \texttt{MCMCDiagnosticTools.jl}~\cite{turing_jl}. Values of $\hat{R}$ close to 1.0 indicate good convergence and well‑mixed chains~\cite{vehtari2021rank}. The ESS assesses and diagnoses poor convergence in the bulk of the distribution due to trends or different locations of the chains. For a given estimand, it is recommended that ESS should exceed $100 \cdot \text{nchains}$~\cite{vehtari2021rank}. Together, $\hat{R}$ and ESS provide complementary diagnostics of chain convergence and posterior exploration quality.

For CNF, posterior approximations were obtained via neural posterior estimation with learned summary statistics~\cite{radev2020bayesflow}, using the BayesFlow software (v1) for amortized Bayesian workflows~\cite{radev2023bayesflow}. We trained an independent neural network for each experiment rather than a single globally amortized model. This preserves amortization over latent trajectories and parameter values within each experiment. Global amortization across experiments is possible in principle but doing so over continuous nuisance variables requires careful conditioning to remain reliable, and broadening the amortization scope was not necessary for our comparison. We therefore adopted per-experiment amortization, which isolates the methodological comparison from the additional design choices that broad amortization would introduce. The resulting per-experiment amortization still enables efficient SBC diagnostics and fast posterior evaluation once training is complete.
Rather than fine‑tuning each method for every individual experiment, which would confound performance differences with tuning effort, we selected CNF hyperparameters through a principled search procedure that we repeated to assess stability. We verified that training was well‑behaved by inspecting training and validation loss curves, which consistently stabilized well before 100 epochs.
The dimension of the learned summary statistics was set to $2n_\theta + 2$, where $n_\theta$ denotes the number of model parameters. Training was conducted offline with 100,000 simulations for the training dataset and 400 simulations for validation. Models were trained for 100 epochs with a batch size of 32. Summary statistics were extracted using a SequenceNetwork~\cite{radev2021outbreakflow} with 64 LSTM units, and the conditional normalizing flow was modeled using an InvertibleNetwork with spline coupling and eight coupling layers, both implemented within the BayesFlow framework. Detailed configurations are provided in Supplementary Information~\nameref{supp:file1}. Handling non‑equidistant time points and missing data required a specialized strategy. We adopted the approach of~\cite{wang2024missing}, which augments the input with binary indicators marking the presence or absence of each entry. This encoding allows the neural network to account for missing values during inference and provides a consistent representation when combining observation functions recorded at different time points. In our setup, only the dimensional consistency of the two observation functions is strictly required; however, we included the binary mask deliberately to facilitate rapid amortization for practitioners who may wish to extend our framework. For the present experiments, this amortization capability was not used: for sparse datasets, the neural networks were trained solely on the specific missingness pattern present in the simulated data to ensure a fair comparison between PF and CNF and to avoid confounding method differences with amortization artifacts. For experiments with dense, equidistant data, this additional encoding was not applied. 
Empirical cumulative distribution function (ECDF) plots were obtained to assess calibration, and parameter recovery diagnostics were used to evaluate the correspondence between inferred and true parameter values. These diagnostics provide insight into whether the amortized posterior captures uncertainty appropriately and whether point estimates remain consistent with the generative parameters.

For the comparison of both methods we used visual and quantitative measures. The visual comparison using density plots of the marginal posteriors allows comparison of the shape and variability of the posterior approximations. This is complemented by a visual comparison of the model fit. For the evaluations we draw 10,000 parameter vector samples from each approximate posterior distribution and, for each parameter vector, simulate an observable trajectory to create an ensemble of independent posterior predictive samples obtained by evaluating $p(y \mid \theta)$ for draws $\theta \sim p(\theta \mid \mathcal{D}^*)$. We then plot the central 50\% and 95\% intervals of the ensemble against the data. As an additional qualitative indicator, we report the maximum a‑posteriori (MAP) estimate computed via Gaussian kernel density estimation using Silverman’s rule~\cite{silverman1998density}, which provides a consistent definition across methods.

For a quantitative measure of inference accuracy, we first assess marginal coverage by reporting how frequently the true generative parameters fall within the 95\% marginal posterior intervals across datasets. We also quantify posterior differences using the 1‑Wasserstein distance in log‑parameter space. All Wasserstein distances are obtained by repeatedly subsampling 1,000 draws per posterior, and we report in‑distribution values to characterize the Monte Carlo error of this procedure. Heuristically, the 1‑Wasserstein distance in log‑parameter space measures the average multiplicative shift required to transform one posterior into another, providing an interpretable notion of scale‑adjusted discrepancy. Concretely, we compute these distances in three settings:
\begin{itemize}
    \item \textbf{Baseline comparisons:} when a baseline is available, we measure the Wasserstein distance between each method and the baseline and compare it to the corresponding in‑distribution reference.
    \item \textbf{Model‑parameterization comparison:} for the full and reparametrized SEIR2v models, we evaluate how the Wasserstein distance changes when keeping the dataset fixed, again reporting the in‑distribution values as a baseline.
    \item\textbf{Within‑method variability:} for one representative parameter set, we assess the variability of each method by computing Wasserstein distances between multiple PF runs or multiple CNF trainings.
\end{itemize}
 Finally, we assess predictive accuracy via the energy score of the posterior predictive distributions. For synthetic experiments, the predictive energy score was also computed with respect to the true parameters used to generate the data. 
For the real-data experiment, true parameters are not available. Instead, we report the energy score relative to published parameter estimates from \cite{ethiopian_data}, which serve as a reference baseline.

All experiments were conducted on ten cores of an AMD EPYC 7f72 3.2 GHz processor with a total of 200 GB of RAM made available to the process. In order to ensure reusability and reproducibility, we made the code and all artificial data used for the experiments available at Zenodo (\url{https://zenodo.org/records/17779579}) and in a GitHub repository (\url{https://github.com/vwiela/Inference-Methods-for-Stochastic-Compartmental-Models.git}).

\section*{Results}\label{section: results}
We examined inference with CNF and PF for stochastic epidemic models through a sequence of in-silico experiments on synthetic data, following the workflow in \autoref{Fig1}(d). To assess the performance of both inference methods against a baseline posterior distribution, we use the standard SIS (\autoref{subsec: sis_model}) and SIR (\autoref{subsec: sir_model}) model with a synthetic prior. The baseline posterior was computed on a finely discretized version of the SDE using a state-of-the-art Hamiltonian Monte Carlo sampler~\cite{hoffman2014no}(Details in Supplementary Information ~\nameref{supp:file1}). We then turn to the complex two-variant SEIR model (\autoref{subsec: seir_model}) with literature-informed priors~\cite{mcaloon2020incubation, fang2020comparisons}, designed to probe the influence of parameter non-identifiability. For this model the computation of a reference posterior is not possible and we examine inference quality in terms of data fit and assess the differences of both methods. Within the two-variant SEIR setting, we additionally study the effects of data sparsity and timing mismatch and investigate a reparametrization aimed at reducing non-identifiability. Finally, we apply both approaches to longitudinal data from Ethiopia~\cite{ethiopian_data} to assess performance under realistic conditions. Evaluation metrics and synthetic dataset generation together with model setups are described in detail in Supplementary Information~\nameref{supp:file1}.


\subsection*{CNF and PF agree with a reference posterior for the SIS model}
\label{subsec:results_sis}
We evaluated convergence and posterior estimates for the SIS model using CNF, PF, and a Hamiltonian Monte Carlo (HMC) baseline applied to a finely discretized version of the SDE. Across 10 synthetic dense datasets, all methods achieved excellent marginal coverage: the true parameters fell within the 95\% credible interval for every posterior marginal (100\% marginal coverage) and produced highly consistent Wasserstein distances. Throughout, each value pair gives the range (minimum to maximum) of the 1-Wasserstein distance in log-parameter space over subsampling repetitions, reflecting the Monte Carlo error of the procedure. Two datasets stood out for distinct reasons, identified by the in-distribution baseline (HMC-HMC), i.e. the distance of an HMC posterior to an independent HMC run on the same dataset. For sis-4, this baseline was itself elevated (0.116-0.214 vs. 0.056-0.119 across the remaining datasets), indicating an intrinsically dispersed posterior rather than a disagreement between methods. We therefore report it separately, as it inflates the between-method aggregate without reflecting method differences. For sis-2, the baseline was at the normal level (0.058-0.081) while only the HMC comparisons rose (0.175-0.199 for CNF-HMC and 0.204-0.230 for PF-HMC) and CNF-PF remained low (0.097-0.137), indicating a localized disagreement against HMC. We report aggregate between-method distances over the remaining datasets; for all cases, visual posterior differences remained minor.

\begin{figure}[tbp!]
    \centering
    \includegraphics[width=.72\linewidth]{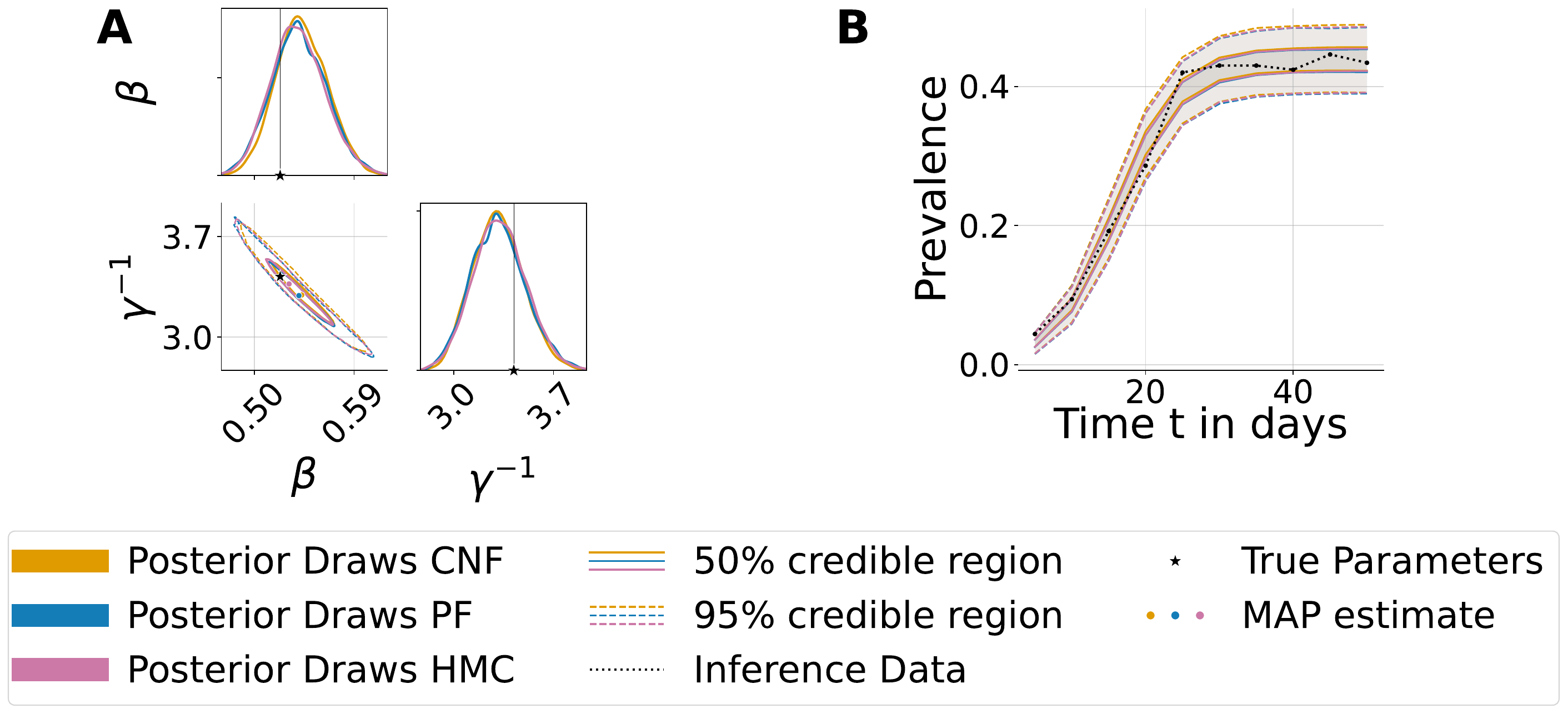}
    \caption{\textbf{Evaluation of the posterior approximations for the SIS model}\\\textbf{A} Posterior approximations from 10,000 samples. Contour gives the 50\% (solid) and 95\% (dashed) credible regions, coloured by method. Diagonals show the 1D marginals. Black stars mark the true parameters, coloured circles the joint MAP estimates. \textbf{B} Posterior predictive fit: bands give the 50\% and 95\% pointwise predictive intervals from the same samples (line styles as in \textbf{A}) with inference data shown as a dotted line.}
    \label{Fig2}
\end{figure}

Across the eight retained datasets, the in-distribution Monte Carlo baseline (HMC-HMC) was small, with values ranging from 0.056 to 0.119. Between‑method distances were only moderately larger: CNF–PF ranged from 0.063 to 0.156, CNF–HMC ranged from 0.065 to 0.176 and PF–HMC ranged from 0.058 to 0.124. Interpreted on the log‑parameter scale, these values correspond to average multiplicative shifts of roughly $6\%-13\%$ for the baseline and $6\%-17\%$, $7\%-19\%$, and $6\%-13\%$ for the between‑method comparisons, respectively. These results indicate that systematic differences between inference algorithms are small, consistent with the strong visual overlap in marginal and joint posteriors (\autoref{Fig2}A). Posterior predictive simulations from all three methods reproduced the observed epidemic trajectories with comparable accuracy (\autoref{Fig2}B), as evidenced by consistent energy scores computed from the posterior predictive distributions of the methods and the ground truth (\autoref{tab: energy sis}).

In terms of sampling behavior, PF showed excellent convergence, with ESS values above 2000 and $\mathbf{\hat{R}}$ near 1. CNF showed good calibration in simulation‑based calibration diagnostics, with near‑uniform rank histograms and accurate parameter recovery. Overall, both PF and CNF matched the predictive and inferential performance of the HMC reference.

Full diagnostics, including ESS, $\mathbf{\hat{R}}$, Wasserstein distance results, calibration plots, and parameter-recovery analyses are provided in Supplementary Information~\nameref{supp:file2}.

\begin{table}[ht]
    \centering
    \captionsetup{width=\linewidth, justification=centering}
    \caption{\textbf{Predictive Energy Score for the SIS model.}}
    \begin{tabular}{l S S S S S}
    \toprule
     & \textit{sis-1} & \textit{sis-2} & \textit{sis-3} & \textit{sis-4} & \textit{sis-5}   \\
     \midrule
    True	&0.034	&0.051	&0.028	&0.019	&0.024 \\
    CNF	&0.035	&0.046	&0.025	&0.018	&0.024 \\
    PF	&0.034	&0.046	&0.025	&0.018	&0.023 \\
    HMC &0.035	&0.046	&0.025	&0.018	&0.023 \\
    \midrule
    & \textit{sis-6} & \textit{sis-7} & \textit{sis-8} & \textit{sis-9} & \textit{sis-10} \\
    \midrule
    True &0.041	&0.025	&0.045	&0.042	&0.041\\
    CNF	&0.040	&0.027	&0.044	&0.035	&0.040\\
    PF	&0.037	&0.022	&0.042	&0.031	&0.040\\
    HMC	&0.037	&0.022	&0.042	&0.031	&0.039\\  
    \bottomrule
    \end{tabular}
    \label{tab: energy sis}
\end{table}

\subsection*{CNF And PF provide accurate estimates for the SIR model.}
\label{subsec:results_sir}
We compared convergence and posterior estimates obtained with CNF and PF on the SIR model, using twelve distinct synthetic dense datasets. As a reference, we additionally computed a reference posterior using Hamiltonian Monte Carlo (HMC) applied to a finely discretized version of the underlying SDE.

\begin{figure}[tbp!]
        \centering
        \includegraphics[width=.72\linewidth]{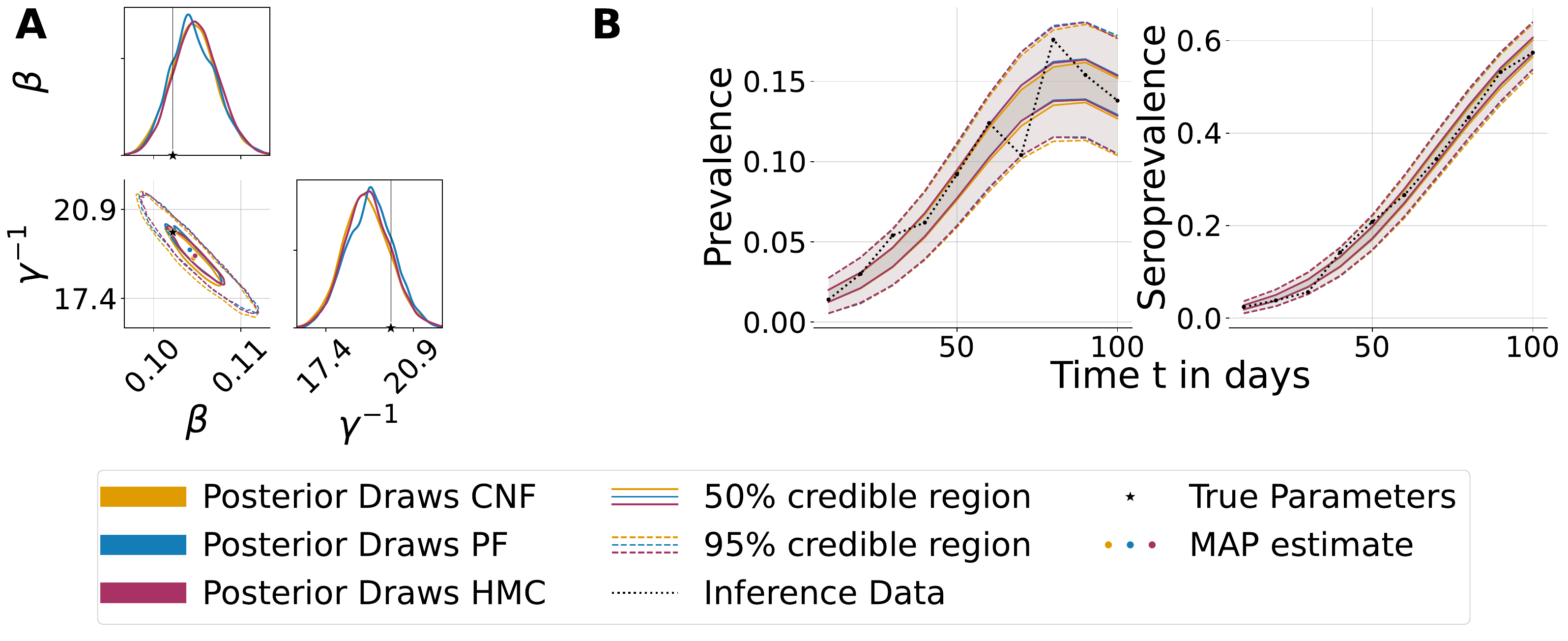}
        \caption{\textbf{Evaluation of the posterior approximations for the SIR model} \newline \textbf{A} Posterior approximations from 10,000 samples. Contour gives the 50\% (solid) and 95\% (dashed) credible regions, coloured by method. Diagonals show the 1D marginals. Black stars mark the true parameters, coloured circles the joint MAP estimates. \textbf{B} Posterior predictive fit: bands give the 50\% and 95\% pointwise predictive intervals from the same samples (line styles as in \textbf{A}) with inference data shown as a dotted line.}
        \label{Fig3}
    \end{figure}

Across all datasets, CNF and PF achieved 83.33\% coverage of the 95\% credible interval and HMC achieved 87.5\% coverage. The three posterior approximations showed high agreement overall: marginal and joint distributions overlapped closely (\autoref{Fig3}A), with CNF exhibiting a small systematic offset relative to HMC and PF. This offset is consistent with the mild posterior miscalibration discussed below. The Wasserstein distances reflect this pattern. The in‑distribution Monte Carlo baseline (HMC–HMC) ranged from 0.034 to 0.096. Between‑method distances were modestly larger: CNF–PF ranged from 0.063 to 0.165, CNF–HMC from 0.076 to 0.183 and PF–HMC ranged from 0.048 to 0.125. Interpreted on the log‑parameter scale, these values correspond to average multiplicative shifts of approximately 3–10\% for the baseline and 6–18\%, 8–20\%, and 5–13\% for the between‑method comparisons, respectively. Detailed results are provided in Supplementary Information~\nameref{supp:file3}. Posterior predictive checks showed that epidemic trajectories generated from all three methods fit the observed data equally well (\autoref{Fig3}B), as evidenced by the close agreement in predictive energy scores (\autoref{tab: energy sir}).

Beyond these direct comparisons, the methods differed in sampling quality. PF achieved excellent convergence, with ESS exceeding 2000 and $\mathbf{\hat{R}}$ values close to 1, indicating robust posterior exploration and well-mixed chains (Supplementary Information \nameref{supp:file3}, Tables S3.2 and S3.3). 
CNF, in contrast, showed deviations in SBC diagnostics (Supplementary Information~\nameref{supp:file3}, Fig. S3.14), reflecting limitations of the amortized posterior. For $\beta$, this miscalibration is largely consistent across all 12 datasets with SBC histograms predominantly sloping upwards. For $\gamma^{-1}$, the histograms did not show one consistent shape. The consistency of the $\beta$ bias points to a structural rather than dataset-specific cause, likely rooted in the constrained joint prior, though we do not attempt a full diagnosis here. This miscalibration geometry matches the observed offset of CNF relative to PF and HMC: the upward-sloping $\beta$ histogram corresponds to a left-shift in the $\beta$ marginal.

Notably, these calibration issues did not carry over to parameter recovery, where recovered values aligned closely with the ground truth and error bars remained narrow (Supplementary Information~\nameref{supp:file3}, Fig. S3.16). Nonetheless, since the posterior is the primary inferential target, we regard this calibration gap as a genuine limitation of the amortized CNF here rather than a purely formal one, even though its impact on point estimates and predictive accuracy is minimal.

In summary, CNF and PF both provide robust Bayesian inference for small stochastic compartmental models. Across datasets, both methods produced equivalent posterior distributions and predictive performance. Notably, CNF, despite being an amortized approach, achieved greater computational efficiency in our experiments (roughly 10-fold) while PF excelled in sampling diagnostics. Moreover, since CNF relies purely on simulations it has consistent computational costs across different parameter vectors. In contrast, PF execution times vary greatly, depending on the posterior geometry. Additionally, the evaluation of the likelihood can be computationally expensive in regions where outcomes are highly sensitive to parameter changes or when parameter vectors do not fit the data well. 

Full diagnostics, including ESS, $\mathbf{\hat{R}}$, Wasserstein distance results, calibration plots, and parameter-recovery analyses are provided in Supplementary Information~\nameref{supp:file3}.

\begin{table}[ht]
    \centering
    \captionsetup{width=\linewidth, justification=centering}
    \caption{\textbf{Predictive Energy Score for the SIR model.}}
    \begin{tabular}{l S S S S S S}
    \toprule
     & \textit{sir-1} & \textit{sir-2} & \textit{sir-3} & \textit{sir-4} & \textit{sir-5} & \textit{sir-6}  \\
    \midrule
    True  & 0.046 & 0.057 & 0.032 & 0.020 & 0.052 & 0.021 \\
    CNF & 0.046 & 0.040 & 0.033 & 0.033 & 0.052 & 0.018  \\
    PF & 0.045 & 0.039 & 0.033 & 0.025 & 0.054 & 0.020  \\ 
    HMC & 0.045 & 0.039 & 0.032 & 0.024 & 0.054 & 0.021  \\ 
    \midrule
    &  \textit{sir-7} & \textit{sir-8} & \textit{sir-9} & \textit{sir-10} & \textit{sir-11} & \textit{sir-12} \\
    \midrule
    True  & 0.050 & 0.054 & 0.056 & 0.071 & 0.061 & 0.073   \\
    CNF & 0.048 & 0.053 & 0.057 & 0.069 & 0.061 & 0.066 \\
    PF & 0.047 & 0.052 & 0.057 & 0.070 & 0.062 & 0.060  \\ 
    HMC & 0.047 & 0.053 & 0.057 & 0.069 & 0.061 & 0.060  \\ 
    \bottomrule
    \end{tabular}
    \label{tab: energy sir}
\end{table}

\subsection*{Non-identifiability issues lead to posterior shape differences}\label{subsec:results_seir2v}
Previous studies have shown that parameters in common epidemiological models are often weakly determined by commonly available datasets~\cite{raimundez2021covid}. To assess how such non-identifiabilities affect inference reliability, we compared PF and CNF on the introduced two‑variant SEIR-model with pronounced non-identifiability for specific parameter combinations. 

\begin{figure}[tbp!]
    \centering
    \includegraphics[width=.6\linewidth]{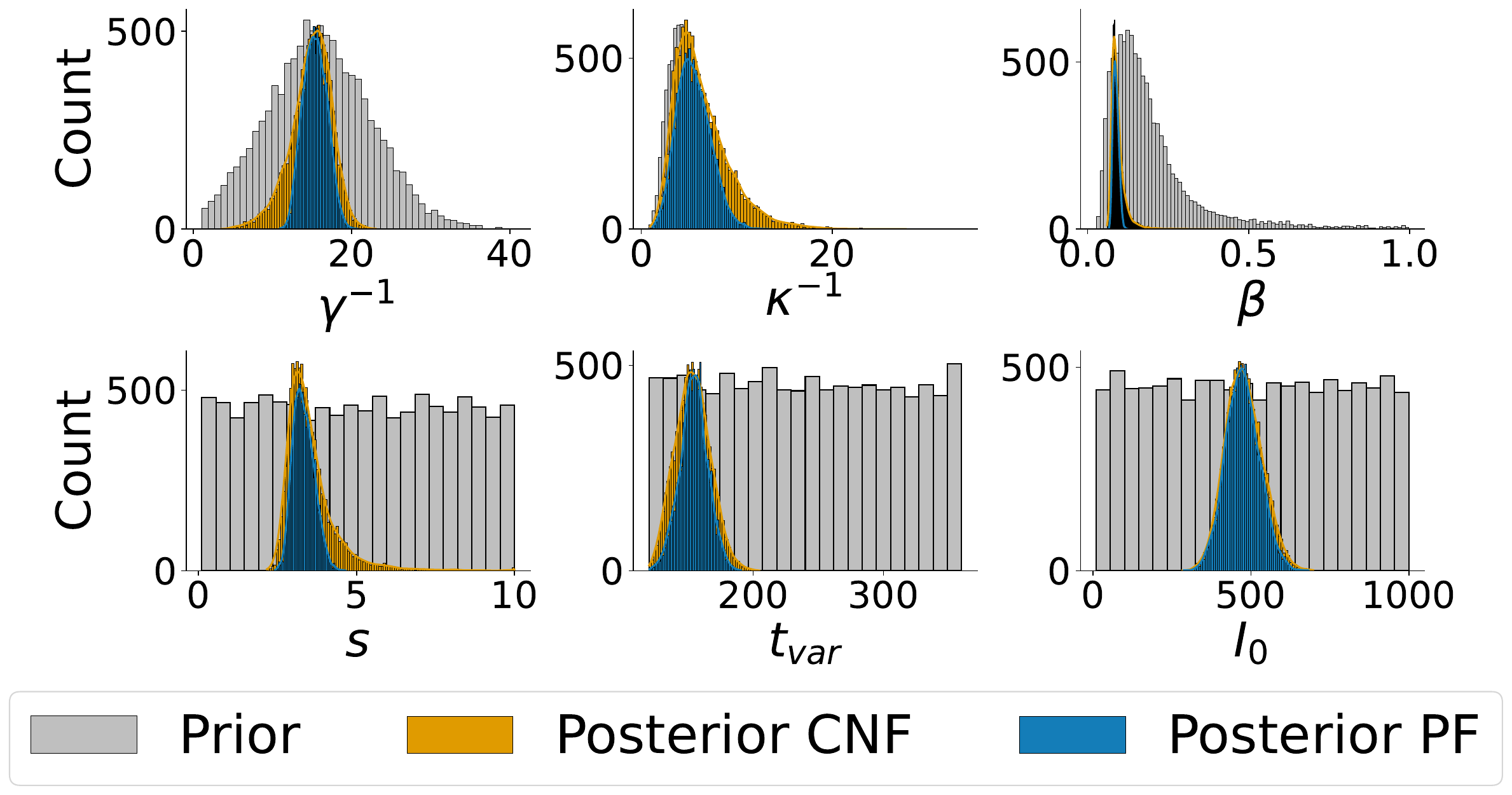}
\captionof{figure}{\textbf{Marginal posterior approximations and prior distribution.} Histograms of posterior approximations and prior distribution for the two-variant SEIR model using a dense dataset.}
\label{Fig4}
\end{figure}

\begin{figure}[tbp!]
        \centering
        \includegraphics[width=0.95\linewidth]{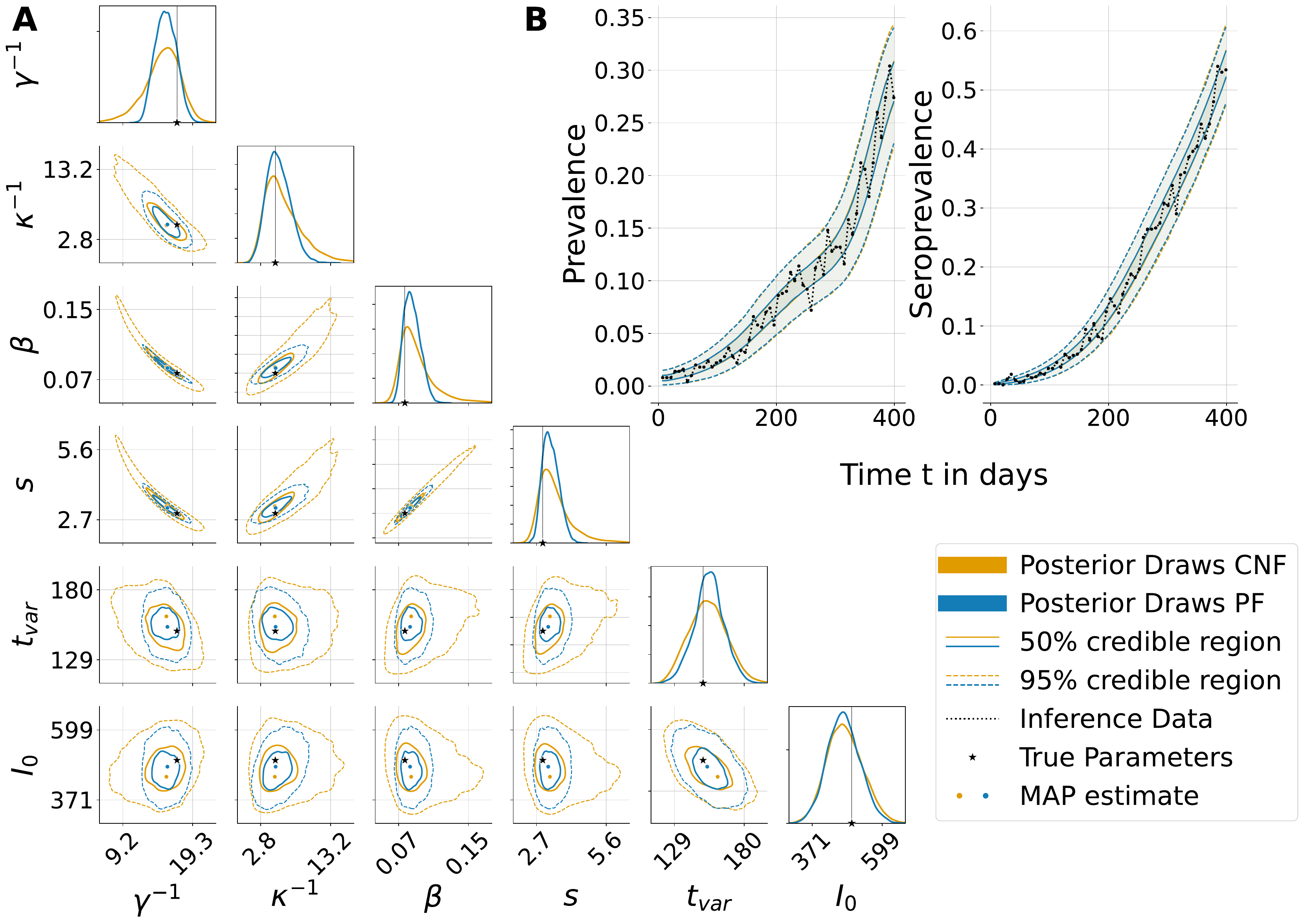}
        \caption{\textbf{Evaluation of the posterior approximations for the two-variant SEIR model with a dense dataset.} \newline \textbf{A} Posterior approximations from 10,000 samples. Contours give the 50\% (solid) and 95\% (dashed) credible regions, coloured by method. Diagonals show the 1D marginals. Black stars mark the true parameters, coloured circles the joint MAP estimates. \textbf{B} Posterior predictive fit: bands give the 50\% and 95\% pointwise predictive intervals from the same samples (line styles as in \textbf{A}) with inference data shown as a dotted line.}
        \label{Fig5}
    \end{figure}

We first examined posterior distributions and predictive performance using twelve dense simulated observational time series. Posterior distributions showed markedly lower variance compared to the corresponding priors, indicating strong information gain from the data (\autoref{Fig4}). The notable exception is $\kappa^{-1}$ under CNF, whose marginal remained close to the prior for some experiments, consistent with weak identifiability~\cite{wang2021posterior}. PF posteriors were generally more concentrated than those of CNF and were largely enveloped by the CNF approximation across parameters (\autoref{Fig5}A). Despite these width differences, both methods generally include the true parameters in their posteriors, with CNF achieving 95.2\% coverage of the 95\% credible interval and PF achieving 91.7\% coverage. Epidemic trajectories generated from both methods fit the data equally well (\autoref{Fig5}B) with comparable energy scores (\autoref{tab: energy seir full dense}). PF produced marginally lower energy scores, highlighting the method’s focus on trajectory agreement through resampling.

The CNF–PF distances were broadly comparable to the variability across repeated CNF runs, suggesting that differences between the two methods remain close to the stochastic variability inherent in CNF. CNF showed good calibration in ECDF plots, though parameter recovery was imperfect for several parameters and markedly worse for $\kappa^{-1}$, which showed weak correlation with the ground truth. PF chains displayed good convergence, with $\mathbf{\hat{R}}$ being close to one across parameters. 
However, ESS were relatively low ($<1000$) for several parameter sets, indicating limited exploration of posterior tails.

\begin{table}[ht]
    \centering
    \captionsetup{width=\linewidth, justification=centering}
    \caption{\textbf{Predictive Energy Score for the full SEIR2V model with dense data.}}
    \begin{tabular}{l S S S S S S S}
    \toprule
     & \textit{d-1-1} & \textit{d-1-2} & \textit{d-2-1} & \textit{d-2-2} & \textit{d-3} & \textit{d-4} & \textit{d-5} \\
     \midrule
    True&	0.113&	0.115&	0.090&	0.112&	0.059&	0.126&	0.089 \\
    CNF&	0.108&	0.106&	0.090&	0.098&	0.058&	0.124&	0.060\\
    PF&	0.107&	0.104&	0.088&	0.098&	0.057&	0.123&	0.059\\
    \midrule
    & \textit{d-6} & \textit{d-7} & \textit{d-8} & \textit{d-9} & \textit{d-10} & \textit{d-11} & \textit{d-12} \\
    \midrule
    True&	0.118&	0.100&	0.101&	0.108&	0.105&	0.099&	0.120\\
    CNF&	0.116&	0.109&	0.087&	0.102&	0.103&	0.088&	0.118\\
    PF&	0.115&	0.114&	0.088&	0.101&	0.097&	0.086&	0.117\\
    \bottomrule
    \end{tabular}
    \label{tab: energy seir full dense}
\end{table}

In practice, observations are often non‑equidistant or partially missing. To reflect this, we conducted additional experiments with reduced observation density (\autoref{subsec: observation model}). 
Results for sparse datasets were consistent with those obtained under dense observations. Posterior distributions from CNF and PF showed similar shapes, and epidemic trajectories generated from both methods fit the data equally well, with energy scores of comparable magnitude (\autoref{tab: energy seir full sparse}). Additionally, CNF remained well calibrated and PF continued to show well‑mixed chains, though ESS values remained low for several parameters, mirroring the diagnostics observed in the dense case. Together, these results indicate that both methods can accommodate sparse and irregular data without loss of inference quality, though CNF requires careful amortization design to handle heterogeneous observation schedules.

\begin{table}[ht]
    \centering
    \captionsetup{width=\linewidth, justification=centering}
    \caption{\textbf{Predictive Energy Score for the full SEIR2V model with sparse data.}}
    \begin{tabular}{l S S S S S S S S}
    \toprule
     & \textit{s-1-1-1} & \textit{s-1-1-2} & \textit{s-1-1-3} & \textit{s-1-2-1} & \textit{s-1-2-2} & \textit{s-1-2-3} &\textit{s-3} &\textit{s-4} \\
     \midrule
     True&	0.042&	0.039&	0.039&	0.041&	0.053&	0.038&	0.018&	0.056\\
    CNF&	0.039&	0.040&	0.035&	0.041&	0.036&	0.033&	0.015&	0.038\\
    PF&	0.037&	0.038&	0.032&	0.033&	0.035&	0.033&	0.015&	0.038\\
    \midrule
    &\textit{s-5}&\textit{s-6}&\textit{s-7}&\textit{s-8}&\textit{s-9}&\textit{s-10}&\textit{s-11}&\textit{s-12}\\
    \midrule
    True&	0.030&	0.037&	0.029&	0.035&	0.052&	0.036&	0.036&	0.042\\
    CNF&	0.017&	0.032&	0.030&	0.029&	0.041&	0.029&	0.033&	0.037\\
    PF&0.015&	0.032&	0.030&	0.029&	0.038&	0.024&	0.030&	0.036\\

    \bottomrule
    \end{tabular}
    \label{tab: energy seir full sparse}
\end{table}

Taken together, these results show that CNF and PF both provide accurate inference for the two‑variant SEIR model, even under pronounced non‑identifiability. CNF yields broader posterior approximations that better capture low‑probability regions, but do not strongly support specific parameter values. PF, in contrast, produces narrower posteriors and more certain point estimates, despite limited exploration of posterior tails due to low ESS. Predictive performance was high for both methods across dense and sparse datasets, with consistent MAP and RMSE values predictive energy scores. Sparse observations introduced implementation overhead for CNFs but did not substantially affect inference quality for either method.
Additionally, CNF are computationally more efficient and can be parallelized easily, while PF are restricted by their sequential nature and only one core per chain can be leveraged. Therefore, CNF were approximately 10-times faster, when parallelized to 20-cores compared to PF on 4-cores, independent of the used dataset (Supplementary Table Information~\nameref{supp:file1}, Table S1.22). 

Full diagnostics, including calibration plots, convergence statistics, Wasserstein-distances and parameter‑recovery analyses, are provided in Supplementary Information~\nameref{supp:file4} for the dense datasets and Supplementary Information~\nameref{supp:file5} for the sparse datasets.

\subsection*{Reducing parameter dependencies improves posterior alignment}\label{subsec:results_reparam} 
Complementing the analysis of the full SEIR2V model, we evaluated the reparametrized version of this model to explore how reduced parameter dependencies affect inference results for PF and CNF. To this end, we analysed twelve synthetic datasets in both the full and the reparametrized settings. In the full SEIR2V model setting, two datasets (\textit{r-2} and \textit{r-5}) did not yield converged Markov chains and one dataset (\textit{r-6}) showed a high $\hat{R}$ value and small ESS for $\beta$ and $\gamma^{-1}$ under our PMMH workflow. This shows the challenges to achieve fast convergence and good mixing in MCMC based methods, if parameters are highly correlated due to unidentifiabilities in the model specification.

\begin{figure}[tbp!]
    \centering
    \includegraphics[width=.95\linewidth]{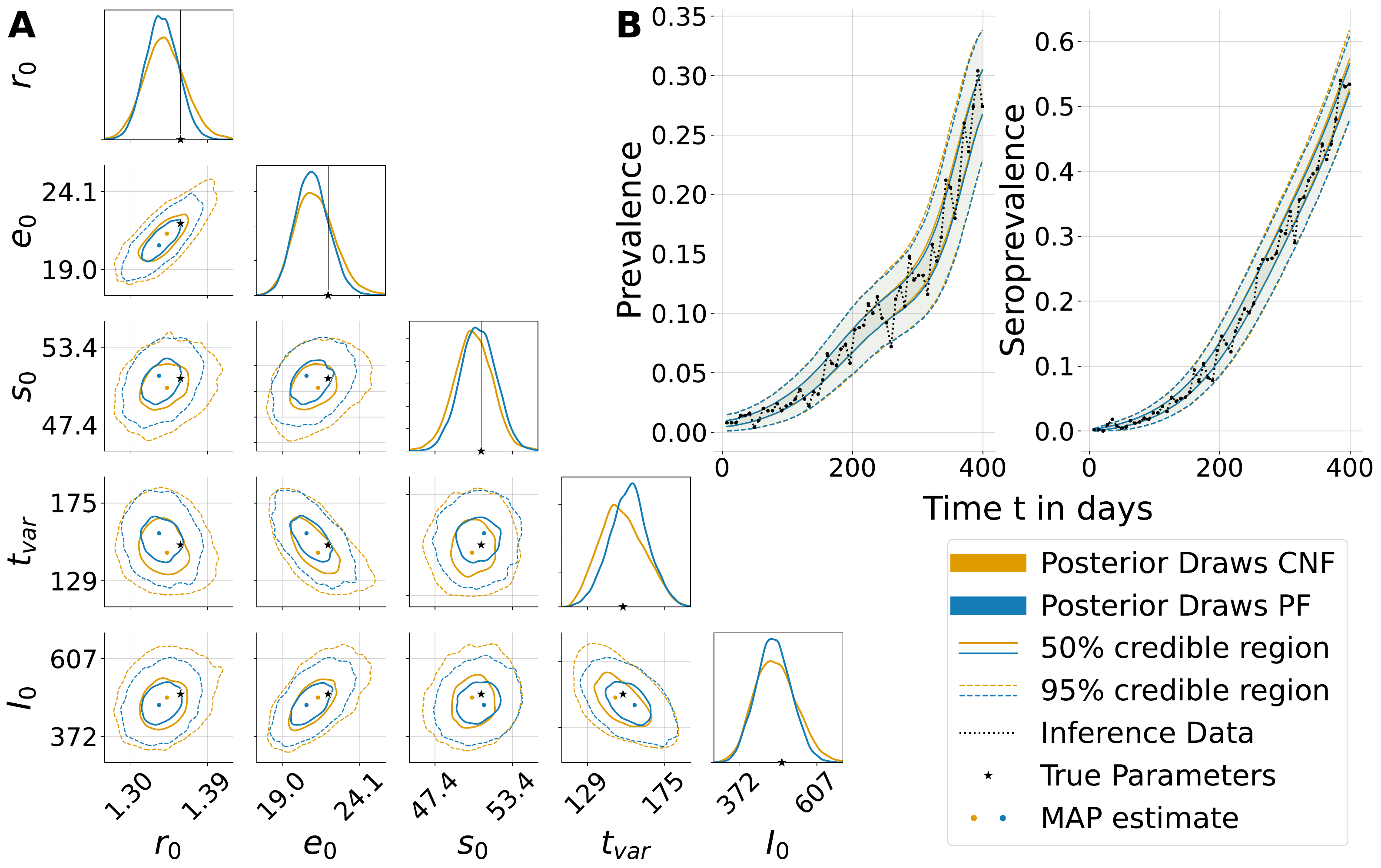}   
    \caption{\textbf{Evaluation of the posterior approximations for the reparametrized two-variant SEIR model with dense data.} \newline \textbf{A} Posterior approximations from 10,000 samples. Contours give the 50\% (solid) and 95\% (dashed) credible regions, coloured by method. Diagonals show the 1D marginals. Black stars mark the true parameters, coloured circles the joint MAP estimates. \textbf{B} Posterior predictive fit: bands give the 50\% and 95\% pointwise predictive intervals from the same samples (line styles as in \textbf{A}) with inference data shown as a dotted line.}
    \label{Fig6}
\end{figure}

In the reparametrized setting, posterior distributions from CNF and PF showed close alignment across parameters, indicating similar exploration of the parameter space (\autoref{Fig6}A). Unlike in the full SEIR2V model, the posterior approximations exhibited no signs of long‑tailed behaviour. This is reflected in the 1‑Wasserstein distances in log‑parameter space, which were numerically smaller under reparametrization and decreased more strongly than the within‑method baselines (CNF–CNF and PF–PF). Moreover, CNF–PF distances were of the same order as the variability observed across independent runs of each method, indicating improved cross‑method alignment. Epidemic trajectories fit the data equally well (\autoref{Fig6}B) with energy scores between CNF, PF, and the true parameters being close to each other in both model settings (\autoref{tab: energy seir reparam dense}, \autoref{tab: energy seir comp dense}).

Additionally, we evaluated calibration and convergence diagnostics. While results in the full model setting were consistent with the ones presented above, CNF showed SBC miscalibration in ECDF plots in the reparametrized setting(Supplementary Information~\nameref{supp:file6}, Fig. S6.14), aligning with the results reported for the identifiable SIR model and hinting at a potential offset in the posterior. PF, in contrast, achieved ESS exceeding 2000 (Supplementary Information~\nameref{supp:file6}, Table S6.2), indicating well‑mixed chains and thorough posterior exploration.

\begin{table}[ht]
    \centering
    \captionsetup{width=\linewidth, justification=centering}
    \caption{\textbf{Predictive Energy Score for the reparametrized SEIR2V model with dense data.}}
    \begin{tabular}{l S S S S S S}
    \toprule
     & \textit{d-1-1} & \textit{d-1-2} & \textit{r-1} & \textit{r-2} & \textit{r-3} & \textit{r-4}  \\
     \midrule
    True Param.	&0.1133	&0.1153	&0.0973	&0.1386	&0.0986	&0.0997\\
    CNF	& 0.1098	&0.1066	&0.0832	&0.1176	&0.1003	&0.1035\\
    PF	& 0.1074	&0.1043	&0.0799&0.1163	&0.0935	&0.1052\\
    \midrule
    & \textit{r-5} & \textit{r-6} & \textit{r-7} & \textit{r-8} & \textit{r-9} & \textit{r-10} \\
    \midrule
    True Param.	&0.1455	&0.1206	&0.1463	&0.1226	&0.0913	&0.0628 \\
    CNF	& 0.1264	&0.1166	&0.1418	&0.0980	&0.0818	&0.05803\\
    PF	& 0.1277	&0.1059	&0.1384	&0.0991	&0.0792	&0.0587\\        
    \bottomrule
    \end{tabular}
    \label{tab: energy seir reparam dense}
\end{table}

\begin{table}[ht]
    \centering
    \captionsetup{width=\linewidth, justification=centering}
    \caption{\textbf{Predictive Energy Score for the full SEIR2V model with dense data from the reparametrized setting.}}
    \begin{tabular}{l S S S S S S}
    \toprule
     & \textit{d-1-1} & \textit{d-1-2} & \textit{r-1} & \textit{r-2} & \textit{r-3} & \textit{r-4}  \\
     \midrule
    True Param.	&0.1132	&0.1148	&0.09754	&0.1388	&0.09826	&0.09969 \\
    CNF	&0.108	&0.1067	&0.09269	&0.1227	&0.1004	&0.1063 \\
    PF	&0.1062	&.1035	&0.07961	&\text{--}&0.09438	&0.1005 \\
    \midrule
    & \textit{r-5} & \textit{r-6} & \textit{r-7} & \textit{r-8} & \textit{r-9} & \textit{r-10} \\
    \midrule
    True Param.	&0.1456	&0.1203	&0.146	&0.124	&0.09142	&0.06312 \\
    CNF	&0.1327	&0.1127	&0.1445	&0.09898	&0.08096	&0.05773 \\
    PF	&\text{--}&0.1059	&0.1382	&0.09892	&0.0796	&0.05737 \\
    \bottomrule
    \end{tabular}
    \label{tab: energy seir comp dense}
\end{table}

In summary, removing parameter dependencies by reducing the number of estimated parameters improved identifiability and brought CNF and PF posterior shapes into closer agreement. While CNF exhibited minor calibration issues, PF benefited from substantially improved ESS, supporting robust posterior exploration in this setting. However, the reparametrization did not enhance predictive performance and relies on reducing the parameter space dimension by fixing $\kappa^{-1}$, which should be approached with caution in applied inference. 

Full diagnostics, including Wasserstein distance analyses, calibration plots, convergence statistics, and parameter‑recovery results, are provided in Supplementary Information~\nameref{supp:file6} for the reparametrized model and Supplementary Information~\nameref{supp:file7} for the full model with datasets of the reparametrized model.

\subsection*{CNF and PF yield consistent results on real-world datasets}
\label{subsec:results_ethiopia}
To complement the synthetic experiments, we applied both methods to data from a longitudinal cohort study of the COVID‑19 outbreak in Ethiopia~\cite{ethiopian_data}. Our setup differed from the original study in that we used a stochastic rather than deterministic formulation, offering greater flexibility and better reflecting the inherent randomness in infection dynamics. Due to the unavailability of ground truth, we used the published mean of the parameter samples as a reference.

\begin{figure}[tbp!]
    \centering
    \includegraphics[width=.95\linewidth]{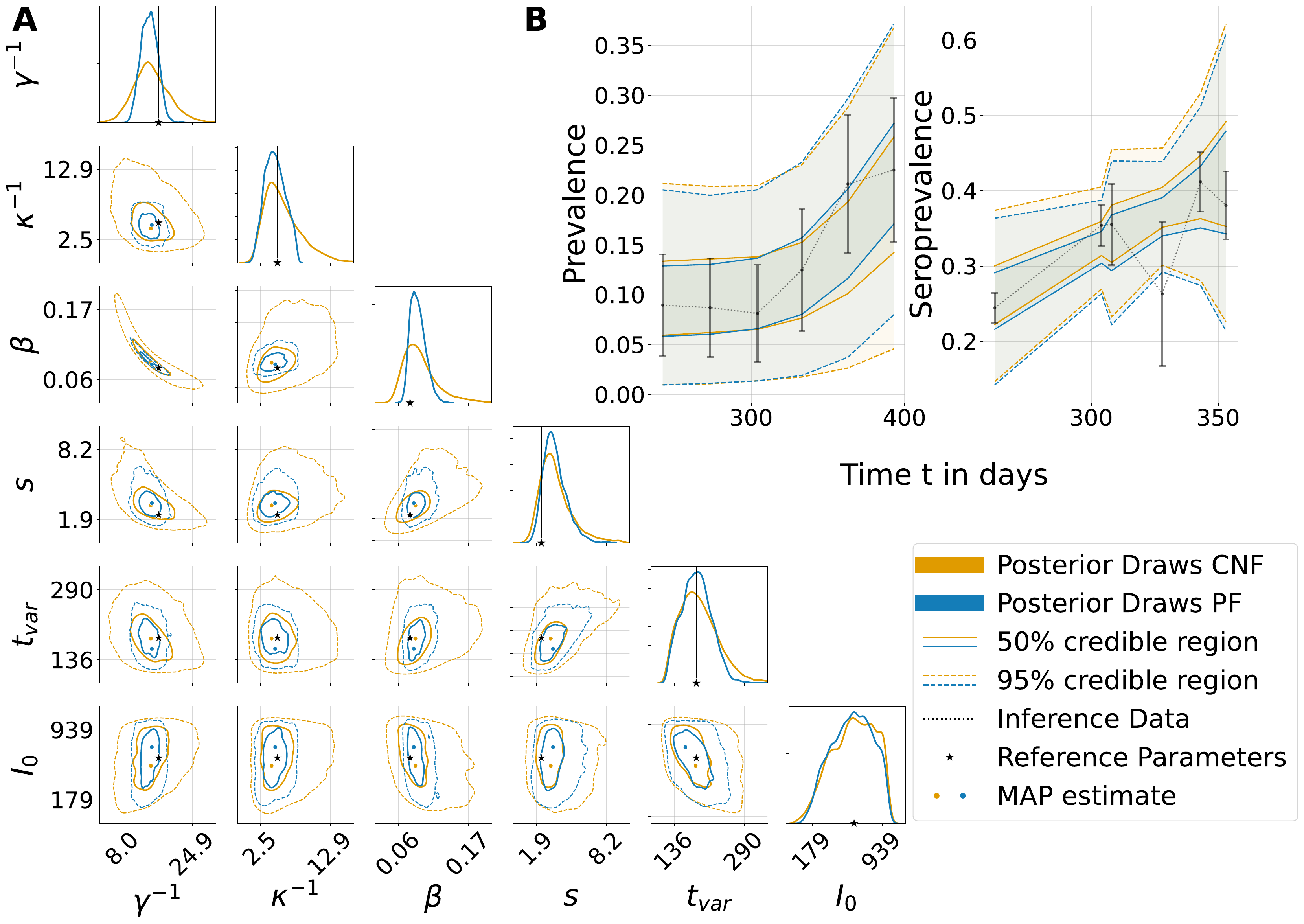}
    \caption{\textbf{Evaluation of the posterior approximations for the two-variant SEIR model with real data.}\\
    \textbf{A} Posterior approximations from 10,000 samples. Contours give the 50\% (solid) and 95\% (dashed) credible regions, coloured by method. Diagonals show the 1D marginals. Black stars mark the reference parameters, coloured circles the joint MAP estimates. \textbf{B} Posterior predictive fit: bands give the 50\% and 95\% pointwise predictive intervals from the same samples (line styles as in \textbf{A}). The inference data (published data from \cite{ethiopian_data}) are shown as a dotted line with error bars.}
    \label{Fig7}
\end{figure}

Posterior approximations from CNF and PF shared a similar overall shape, with the PF posterior largely nested within the broader CNF and exhibiting reduced variance, particularly for the rate parameters $\beta$, $\kappa^{-1}$, and $\gamma^{-1}$ (\autoref{Fig7}A). Trajectory ensembles derived from the inferred posteriors fit the data well, even under the relatively high noise level reported in the original study; one seroprevalence point was not captured by either method within their uncertainty intervals (\autoref{Fig7}B). Predictive energy scores between CNF and PF improved on the energy score calculated from the the originally published parameter set, indicating that both inference methods provide a better probabilistic fit to the observed data than the originally published parameter set (\autoref{tab: energy seir full ethiopia}).

Diagnostics were consistent with patterns observed in controlled experiments: CNF exhibited good calibration alongside imperfect parameter recovery, while PF demonstrated convergence despite low effective sample sizes for $\beta$ and $\gamma^{-1}$.

\begin{table}[ht]
    \centering
    \captionsetup{width=\linewidth, justification=centering}
    \caption{\textbf{Predictive Energy Score for the full SEIR2V model with real-world data.}}
    \begin{tabular}{l c}
    \toprule
     & Ethiopia  \\
     \midrule
     Pub. & 0.131 \\
     CNF & 0.107 \\
     PF & 0.096 \\
    \bottomrule
    \end{tabular}
    \label{tab: energy seir full ethiopia}
\end{table}

Taken together, applying both methods to the Ethiopian cohort data yielded results consistent with the synthetic analyses, underscoring the robustness of each approach. The trade‑offs identified in controlled settings—CNF’s broader posterior coverage versus PF’s tighter estimates with lower ESS—carried over to this real‑world application, while predictive performance remained strong for both methods.

Full diagnostics, including calibration plots, convergence statistics, and parameter‑recovery results, are provided in Supplementary Information~\nameref{supp:file8}.

\section*{Discussion}\label{section: discussion}
Responses to public health crises, such as outbreaks of infectious diseases, create a need for informed decision making by realistically modeling the dynamics using stochastic models. Yet, fitting such models to observed data present many difficulties and several methods were developed to address this challenge.
In this work, we systematically evaluated two state-of-the-art methods for Bayesian parameter inference, Conditional Normalizing Flows (CNF) and pseudo-marginal particle MCMC, referred to as Particle Filters (PF), across a range of stochastic compartmental models. Our analysis encompassed synthetic and real-world data, varied model structure, and different levels of data sparsity, yielding crucial insights into the strengths and limitations of each methodology when applied in epidemiological modeling scenarios. Where feasible, we calculated a reference posterior using Hamiltonian Monte Carlo (HMC) on a finely discretized version of the stochastic models. Across all investigated scenarios, both methods produced accurate fits and informative posteriors. However, this study revealed systematic differences of uncertainty in the parameter space due to different posterior geometries and model design choices. 

In the context of small stochastic models, such as the SIR model and the SIS model, both CNF and PF yielded highly consistent posterior distributions and closely matched predictive epidemic trajectories. The marginal and joint posterior estimates and MAP estimates were near the data-generating parameters and the predictive energy scores confirmed high predictive accuracy for both methods. CNFs exhibited miscalibration for the SIR model, which resulted in small offsets relative to PF and HMC. In contrast to that, CNF exhibited very good recovery plots in these settings. As an exact method, PF showed a good recovery of the true parameters and a high agreement with the reference posterior distributions, that were closely matched by the CNF. Together, this indicates that, once trained, CNF can deliver inference quality on a par with an exact MCMC based baseline while enabling possible amortization across different datasets and yielding high computational efficiency in the inference phase but miscalibration needs to be handled and analysed carefully. 

Expanding the analysis, we applied both methods to a more complex stochastic two-variant SEIR model with practical non-identifiabilities and parameter couplings, that does not admit a reference posterior. In this setting, both methods fit the data well and achieved similar predictive power and uncertainty, yet PF produced narrower marginal posteriors that were largely enclosed by the broader CNF posteriors and simultaneously showed low effective sample sizes for several parameters. This pattern is consistent with resampling in PF concentrating particles in high-likelihood basins, i.e., on the fit to the data. This property can limit tail exploration in the parameter space especially when the posterior surface contains ridges or mountain-pass structures. Additionally, PF based methods are, as most MCMC methods, highly dependent on the choice of the initial parameter value. Uninformative initial parameter values lead to degenerate likelihood estimate and therefore stuck Markov chains. CNF achieved good calibration results in this setting, although its recovery plots indicated that parameter recovery was not perfect. The wider posteriors for CNF contrasts with prior literature where amortized inference often underestimates uncertainty~\cite{hermans2022trust}. 
Despite these differences, both methods captured the inference data accurately and showed strong predictive performance, as evidenced by low and similar predictive energy scores. Their application in this setting shows that amortized methods may struggle with calibration, yet still provide strong predictive performance and exact MCMC based method estimate the bulk of the posterior well while underestimating its width caused by insufficient exploration of the parameter space.

Using sparsely sampled datasets with different missingness patterns did not change the qualitative estimation results of both methods. But the two methods inherently treat sparse datasets differently. While CNF methods can accommodate such dataset variability, amortizing over all relevant combinations of time points may become computationally expensive and risks unreliable extrapolation if not carefully designed. This challenge is specific to setups that aim for broad amortization; alternatively, one can train CNFs for fixed missingness patterns. In contrast, PF methods naturally handle non-equidistant and partially missing data via their sequential simulation and weighting scheme, which decomposes the likelihood over time points and observables without requiring amortization or any other adaptation of the method.

Afterwards we introduced a reparametrization of the two-variant SEIR model that combined or transformed parameters based on their observed dependencies. This adjustment resulted in better alignment between CNF and PF posterior estimates and markedly increased PF convergence and mixing. These improvements suggest that the posterior geometry, rather than methods alone, leads to robust inference in the parameter space and model choices should be investigated as thoroughly as tuning of methods. CNF achieved worse calibration than in the full model case but its recovery plots demonstrated good parameter recovery. Both methods showed robustness to practical non-identifiabilities in their quality of fit and predictive power.

Finally, applying both methods to a real dataset from a longitudinal cohort study in Ethiopia confirmed the simulation study findings and yielded broadly similar posterior approximations and epidemic trajectories. Although obtained MAP estimates differed from previously published parameter values, especially for CNFs, both methods captured the data dynamics and produced model fits with predictive energy scores close to or improving upon the deterministic analysis~\cite{ethiopian_data}. By demonstrating their effectiveness in a real-world setting, our analysis underscores the potential of these methods to inform public health strategies and decision-making processes.

We also compared the computational efficiency of the two methods. To ensure a meaningful comparison, both methods were calibrated to produce the same number of samples from the parameter posterior, with CNF's neural network trained on a CPU. Despite this constraint, CNF is roughly 10 times faster than PF, due to PFs sequential nature and the higher number of simulator calls. The bulk of CNF’s computational time is spent on training, but once trained, inference can be performed in mere seconds. CNF's amortization allows multiple inferences, providing, in this scenario, a significant advantage over PF, which requires separate computations for each run. To achieve greater speed, CNF can make use of a parallelized simulator or training the neural network on a GPU. Meanwhile, PF could benefit from the use of more advanced filtering algorithms such as guided PFs~\cite{snyder_2011, chopin_2020} or using correlated particles~\cite{wiqvist_2021} to avoid degeneracy or too high particle numbers. Switching to a more advanced outer MCMC sampler such as a Gibbs sampler~\cite{chopin2015} or a Metropolis-Adjusted Langevin algorithm~\cite{lowe2023accelerating} and parallelization over particles could speed-up the inference procedure as well. Additionally, the lack of amortization renders assessment of calibration and hyperparameter tuning for PF too computationally expensive.

The scope of this work leads to several limitations. For the more complex two-variant SEIR model no reference posterior was available, leaving the assessment based on comparative observations between the two chosen methods. The observation model employed additive Gaussian noise for clarity and comparability; overdispersed counts or serological adjustments may be better represented by negative‑binomial or beta‑binomial formulations in other settings. Our use of an Euler–Maruyama integration with positivity clipping introduces discretization error whose interaction with inference can lead to numerical instabilities near boundaries or in small populations. Algorithmic choices were guided by the ease of their applicability in praxis: We focused on a bootstrap PF within PMMH and a tuned CNF architecture with consistent hyperparameters and fixed simulation budget across our datasets and experimental setups. Alternative PF proposals (e.g., guided/auxiliary filters or tempering within SMC) and alternative SBI estimators (likelihood or ratio estimation) or more exhaustive hyperparameter-tuning could shift efficiency–accuracy trade‑offs. To avoid out‑of‑distribution failure modes, CNFs were trained per experiment rather than globally amortized across all data regimes, which tempers their wall‑clock advantage in this study; broader amortization remains attractive but demands careful conditioning on meta‑data such as time‑grid irregularity and noise scales~\cite{radev2023bayesflow, wang2024missing}. 

In the SBI literature, recent advances focus on workflows that narrow the gap between neural SBI and Monte Carlo by using learned proposals inside sequential Monte Carlo or MCMC~\cite{papamakarios2019sequential, hermans2020likelihood}. Variational and amortized SMC approaches train proposal families to reduce weight degeneracy and improve effective sample sizes; conversely, particle methods can provide high‑quality samples to bootstrap or regularize neural estimators. Looking forward, future work in this context of hybrid approaches could focus on a flow trained in the spirit of neural posterior estimation that proposes parameter values to a PMMH kernel or guide latent‑state proposals. These methods could combine the tail coverage and speed of neural estimators with the asymptotic correctness and model flexibility of pseudo‑marginal schemes. Recent generative modeling advancements, i.e., flow matching and diffusion‑style transports, offer further avenues for robust posterior learning with stronger inductive biases and improved stability~\cite{lipman2022flow, draxler2024free, arruda2025}.
Additionally, the use of adaptive prior schemes where the prior is updated in light of the available data, through approaches such as empirical Bayes, hierarchical modeling, or iterative refinement, may help to mitigate issues of non-identifiability while improving robustness. Such strategies may enhance performance in particularly challenging applications that use real world data where information is both sparse and noisy.

In summary, our comparison suggests that both CNF and PF are viable inference strategies for complex epidemic models, with different focus on parameter space exploration and data fitting. CNF tends to provide a more extensive posterior, whereas PF typically produces tighter distributions. Moreover, CNF's substantially lower computational cost and amortization capabilities, achieving approximately 10 times faster performance than PF under comparable conditions, makes CNF especially advantageous for large-scale or time-sensitive applications. In contrast, PF maintain exactness (in the pseudo‑marginal sense) under model changes at the cost of potential tail under‑exploration, rendering PF a robust choice across models. The use of sparsely sampled data with different missingness patterns highlights the applicability of both methods and their robustness to different data settings. By clarifying the trade‑offs and giving insights into the performance of these two practical inference methods, this work provides the groundwork for evaluating inference engines ready for uncertainty‑aware epidemic forecasting and real-time support of public-health decisions.

\section*{Supporting information}

\section*{Supporting Information}

\paragraph*{S1 Text}
\label{supp:file1}
\textbf{Supporting Information on methods.}

\paragraph*{S2 Results}
\label{supp:file2}
\textbf{Supplementary Results SIS-Model}

\paragraph*{S3 Results}
\label{supp:file3}
\textbf{Supplementary Results SIR-Model}

\paragraph*{S4 Results}
\label{supp:file4}
\textbf{Supplementary Results SEIR-Model with Dense Data}

\paragraph*{S5 Results}
\label{supp:file5}
\textbf{Supplementary Results SEIR-Model with Sparse Data}

\paragraph*{S6 Results}
\label{supp:file6}
\textbf{Supplementary Results Reparametrized SEIR-Model}

\paragraph*{S7 Results}
\label{supp:file7}
\textbf{Supplementary Results Comparison Full SEIR-Model with Reparametrized Model}

\paragraph*{S8 Results}
\label{supp:file8}
\textbf{Supplementary Results SEIR-Model with Real Data}

\section*{Author Contributions}
V.W.: Conceptualization, Application PF, Analysis PF, Visualization, Writing - original draft, Writing - review \& editing.
N.W.: Conceptualization, Application CNF, Analysis CNF Visualization, Writing - original draft, Writing - review \& editing. 
L.C.: Software PF
M.K.: Supervision, Funding acquisition, Writing - review \& editing
J.H.: Supervision,  Project administration, Funding acquisition, Writing - review \& editing

\section*{Funding}
This work was supported by the Deutsche Forschungsgemeinschaft (DFG, German Research under Germany’s Excellence Strategy (EXC 2151—390873048), by the European Union via ERC grant INTEGRATE (grant no 101126146) to J.H. and by the University of Bonn via the Bonn Center for Mathematical Life Sciences and the Schlegel Professorship of J.H.). This work received further funding from the Initiative and Networking Fund of the Helmholtz Association (grant agreement number KA1-Co-08, Project LOKI-Pandemics). There was no additional external funding received for this study.

\clearpage
\bibliographystyle{plainnat}
\bibliography{literature.bib}

\end{document}